\documentclass[%
reprint,
superscriptaddress,
nofootinbib,
amsmath,
amssymb,
aps,
floatfix,
showkeys,
]{revtex4-2}

\usepackage[colorlinks,citecolor=blue,linkcolor=blue,anchorcolor=blue,filecolor=blue,
urlcolor=blue]{hyperref}

\usepackage{graphicx,color,xcolor}
\usepackage{amsfonts,amsmath}
\usepackage{amssymb,ulem}
\usepackage{dcolumn}
\usepackage{bm,lipsum}
\usepackage[utf8]{inputenc}
\usepackage{subeqnarray}
\usepackage{array}
\usepackage{epsfig,hyperref}
\usepackage{morefloats}
\usepackage{relsize}
\usepackage{url}

\usepackage{comment}

\begin{document}

\preprint{ }

\title{Role of the symmetry energy slope in neutron stars: exploring the model-dependency}

\author{Luiz L. Lopes}

\affiliation{ Centro Federal de Educac\~{a}o Tecnol\'{o}gica de Minas Gerais Campus VIII; CEP 37.022-560, Varginha - MG - Brasil\\}

\date{\today}

\begin{abstract}
Using six different parametrizations of the quantum hadrodynamics (one of which is original), I study how different values of the symmetry energy slope ($L)$ affect some microscopic and macroscopic properties of neutron stars, such as the proton fraction, the maximum mass, the radius of the canonical 1.4$M_\odot$ star and its dimensionless tidal parameter $\Lambda$. 
I show that while most quantities present the same qualitative results, the tidal parameter can increase or decrease with the slope, depending on the model. Moreover, special attention is given to the minimum mass that enables the direct URCA process to occur in neutron stars' interiors ($M_{DU})$. Assuming the weak constraint  $M_{DU}~>~1.35M_\odot$, one can see that the maximum value of $L$ that satisfies it lies between 79 and 86 MeV. A range of only 7 MeV.
Therefore, $M_{DU}$ is an easy way to impose upper bounds to the slope.
\end{abstract}

\maketitle

\section{Introduction}

Neutron stars are among the densest objects in the universe. With masses reaching two solar masses while the radii are not larger than 15 km, their central density can reach several times the density of the atomic nuclei. Although our knowledge about nuclear physics near the saturation point had a great leap in the last decade, we still have a very foggy vision for densities above three or four times this point. Due to this, it is easy to find in the literature models that predict similar values near the saturation density but very different behavior at high densities.

In two different review papers (ref.~\cite{Dutra2014,Micaela2017}), the authors were able to constraint five nuclear quantities at saturation density: the saturation density itself $(n_0)$, 
the nucleon effective mass $(M^{*}_N/M_N)$ the binding energy per baryon $(B/A)$, the 
(in)compressibility ($K$), and the symmetry energy $(S_0)$. A sixth quantity, the symmetry energy slope, or simply the slope ($L$) is still a matter of debate. In the earlier 2010s, most studies pointed to a relatively low value for $L$. For instance, in refs.~\cite{Paar2014,Steiner2014,Lattimer2013}  upper limits of 54.6, 61.9, and 66 MeV respectively were suggested. However, the situation has changed in the last couple of years, and new experiments have pointed to a significantly higher upper limit.
For instance, in a study about the spectra of pions in intermediate energy collisions, an upper limit of 117 MeV was obtained~\cite{pions}, while in one of the PREX II analyses~\cite{PREX2} an upper limit of 143 MeV was suggested.
All these conflicting results have been well summarized in a recent paper~\cite{Tagami2022}: the CREX group points to a slope in the range 0 $<~L~<$ 51 MeV, while PREX II results point to 76 MeV $<~L~<$ 165 MeV. The CREX and PREX II results do not overlap. It is a huge problem that must be solved.

In this work, I study the influence of the symmetry energy slope by fixing the five well-known quantities at the saturation density and varying only the slope, $L$. Moreover, to study the model-dependency of the results I use six different parameterizations of the quantum hadrodynamics (QHD)~\cite{Serot_1992,Miyatsu2013}. I only impose two prior constraints for the models. I require that at least four of the five parameters of the nuclear matter at the saturation point satisfy the constraint coming from ref.~\cite{Dutra2014,Micaela2017}, and that all models predict neutron stars with masses $M~>2.0M_\odot$. This feature is imperative, once the existence of very massive stars is well established, such as the  PSR J0348+0432 with a mass of $2.01 \pm 0.04 M_\odot$~\cite{Antoniadis} and the PSR J0740+6620, with $M = 2.08\pm 0.07~M_\odot$~\cite{Riley2021}.

Only after that do I start to investigate which parametrization and with what values of $L$ can fulfill other constraints coming from nuclear astrophysics. For instance, the radius of the PSR J0740+6620 lies between 11.41 km $<~R~<$ 13.70 km as suggested in ref.~\cite{Riley2021}. Concerning the canonical 1.4 $M_\odot$ star, two NICER teams have pointed to a limit of $13.85 ~\mathrm{km}$ \cite{Riley:2019yda} and $14.26 ~\mathrm{km}$ \cite{Miller:2019cac}. These results were refined in ref.~\cite{Miller2021}  to 11.80 km $<R_{1.4}< 13.10$ km. A more conservative constraint coming from state-of-the-art theoretical results at low and high baryon density points to an upper limit of $R_{1.4}$ $<$ 13.6 km~\cite{Annala2018PRL}. Still on the canonical star, another important quantity and constraint is the so-called dimensionless tidal deformability parameter $\Lambda$. The gravitational wave observations by LIGO/VIRGO in the  GW170817 event put the constraints on the dimensionless tidal parameter of the canonical star  $\Lambda_{1.4}<800$~\cite{Abbott2017}. This result was then refined in ref.~\cite{AbbottPRL}, to  $70<\Lambda_{1.4}.<580$.

I also pay special attention to the possible presence of the direct URCA (DU) process in neutron stars. The so-called standard model of neutron star cooling is
based upon neutrino emission from the interior, which is dominated by the modified URCA process.
However, if the proton fraction exceeds some critical value $(X_{DU})$ in the range $11–15\%$, the process will be dominated by the Direct URCA process, a process one million times more efficient~\cite{Lattimer1991,Lattimer2004,Fatto2012,Dohi2019}. As in beta-stable matter, the condition expected in neutron stars' interior, the proton fraction grows with the density, the more massive the star; the higher will be the proton fraction. For each parametrization and slope value, I calculate the minimum mass that enables the DU process $(M_{DU})$. Moreover, there are also some constraints related to it.  Ref.~\cite{klahn2006}  pointed out that any acceptable equation of state (EoS) shall not allow the direct URCA process to occur in neutron stars with masses below  1.5$M_\odot$. Such constraint is corroborated by a recent study on the statistical theory of thermal evolution of neutron stars, which suggests that the minimum mass that allows the DU process lies between 1.6 - 1.8$M_\odot$~\cite{Yakovlev2015}. These are, nevertheless, faced as strong constraints. A weak constraint is presented in  refs.~\cite{klahn2006,Page1992,YAKOVLEV2004a,Yakovlev2004b} and points that $M_{DU}~>~1.35M_\odot$.

Therefore, for the six parametrizations of the QHD, I fix the five well-known parameters at the saturation density and run over the slope from 44 MeV up to 92 MeV, to rule out values of $L$ that predict $M_{DU}~<~1.35M_\odot$.   Moreover, to keep the symmetry energy fixed while varying the slope, I add the non-linear $\omega-\rho$ coupling as presented in the IUFSU model~\cite{IUFSU,Rafa2011,dex19jpg}. It is also possible to obtain $L$ above 92 MeV with the help of the scalar-isovector $\delta$ meson~\cite{KUBIS1997,Liu2002,Lopes2014BJP}. However, as it will become clear in the text, this is not necessary because $L$ = 92 MeV already produces $M_{DU}~<~1.35M_\odot$. Moreover, two recent papers~\cite{lopescesar,lopes2024PRC} indicate that higher values of the slope are in disagreement with some constraints coming from neutron stars' observations, once it predicts the radii for the canonical stars outside the limits inferred by the NICER observations~\cite{Riley:2019yda,Miller:2019cac}, as well a hadron-quark phase transition very close or even below the saturation density.

\section{Formalism and parametrizations}

The extended version of the QHD~\cite{Serot_1992}, which includes the $\omega\rho$ non-linear coupling~\cite{IUFSU,Rafa2011,dex19jpg}  has the following Lagrangian density in natural units:
\begin{eqnarray}
\mathcal{L}_{QHD} =  \bar{\psi}_N[\gamma^\mu(\mbox{i}\partial_\mu  - g_{\omega}\omega_\mu   - g_{\rho} \frac{1}{2}\vec{\tau} \cdot \vec{\rho}_\mu)+ \nonumber \\
- (M_N - g_{\sigma}\sigma )]\psi_N  -U(\sigma) + \frac{1}{2} m_v^2 \omega_\mu \omega^\mu +   \nonumber   \\
  + \frac{1}{2}(\partial_\mu \sigma \partial^\mu \sigma - m_s^2\sigma^2) + \frac{\xi g_\omega^4}{4}(\omega_\mu\omega^\mu)^2  + \nonumber \\ - \frac{1}{4}\Omega^{\mu \nu}\Omega_{\mu \nu} + \Lambda_{\omega\rho}(g_{\rho}^2 \vec{\rho^\mu} \cdot \vec{\rho_\mu}) (g_{\omega}^2 \omega^\mu \omega_\mu) + \nonumber \\
 + \frac{1}{2} m_\rho^2 \vec{\rho}_\mu \cdot \vec{\rho}^{ \; \mu} - \frac{1}{4}\bf{P}^{\mu \nu} \cdot \bf{P}_{\mu \nu}. \label{s1} 
\end{eqnarray}
 The $\psi_N$  is the  Dirac field of the nucleons.  The $\sigma$, $\omega_\mu$,  and $\vec{\rho}_\mu$ are the mesonic fields.
 The $g's$ are the Yukawa coupling constants that simulate the strong interaction, $M_N$ is the nucleon mass and  $m_s$, $m_v$, and $m_\rho$ are
 the masses of the $\sigma$, $\omega$,  and $\rho$ mesons respectively. The $\xi$ is related to the self-interaction of the $\omega$ meson, while the $\Lambda_{\omega\rho}$ is a non-linear coupling between the $\omega$-$\rho$ mesons and controls the symmetry energy and its slope~\cite{dex19jpg}. 
The $U(\sigma)$ is the self-interaction term introduced in ref.~\cite{Boguta} to fix the compressibility:

\begin{equation}
U(\sigma) =  \frac{\kappa M_N(g_{\sigma} \sigma)^3}{3} + \frac{\lambda(g_{\sigma}\sigma)^4}{4} \label{sbog} .
\end{equation} 

Furthermore, leptons are added as free fermions to account for the chemical stability. The EoS is then obtained in mean field approximation (MFA) by calculating the components of the energy-momentum tensor. The detailed calculation of the EOS in the mean field approximation can be found in ~\cite{Serot_1992,IUFSU,Lopes2014BJP,Lopes2022ApJ,Miyatsu2013,Tolos2017,Glenbook} and the references therein.

I use six different QHD parameterizations to study the model dependency of the slope in neutron stars' properties. Five of them are well-known in the literature. They are (from the softer to the stiffer one) the NL$\rho$~\cite{Liu2002}, the L3$\omega\rho$~\cite{Lopes2022CTP}, the GM1~\cite{GlenPRL}, the FSU2H~\cite{Tolos2017} and the BigApple~\cite{BigApple}. The sixth one is completely original. Recently, ref.~\cite{Dex2024arxiv} reported that 
the NICER group during the April APS meeting indicates that the PSR J0437-4715 has a radius of only $11.36^{+0.95}_{-0.63}$ km. With a mass very close to the canonical star, $M_{J0437}$ =  1.418$M_\odot$, such radius range presents a very strong constraint for the upper limit: $R_{J0437} < 12.31$ km. The original parametrization can fulfill such strong constraint (at least for $L = 44$ MeV) and simultaneously predict a maximum mass above 2.2$M_\odot$. This feature can be potentially important because ref.~\cite{romani} indicates that the black widow pulsars PSR J0952-0607 has a mass  $M = 2.35\pm0.17 M_\odot$. Moreover, the original parametrization (which I call here L1$\omega^4$) fulfills the five constraints at the saturation density presented in ref.~\cite{Dutra2014,Micaela2017}.

For each parametrization, except for the slope, all the other five quantities ($n_0$, $M^{*}_N/M_N$, $B/A$, $K$, $S_0$)  are fixed. This implies that, except for $g_\rho$ and $\Lambda_{\omega\rho}$, all other parameters of the model are also fixed. In Tab.~\ref{TL1}, I present the fixed parameters of the six models, the prediction of the five fixed quantities, and the constraints coming from ref.~\cite{Dutra2014,Micaela2017}. In Tab.~\ref{T2}, I present the values of $g_\rho$ and $\Lambda_{\omega\rho}$ for each model and for four different slope values.

As can be seen from Tab.~\ref{TL1}, four of the six parametrizations satisfy all the five constraints of nuclear matter. The FSU2H fails to fulfill the effective nucleon mass at saturation density, but its value is only $\approx$ $1\%$ below the bottom limit. On the other hand, the GM1 parametrization presents a compressibility around $15\%$ above the upper limit.
I nevertheless kept the GM1 parametrization precisely to study the effects of such a high value of $K$.

\begin{widetext}
\begin{center}
\begin{table}[h]
\begin{center}
\begin{tabular}{c|cccccc|ccc}
\hline
   & NL$\rho$~\cite{Liu2002} & L3$\omega\rho$~\cite{Lopes2022CTP}  & GM1~\cite{GlenPRL} & FSU2H~\cite{Tolos2017}  &BigApple~\cite{BigApple} & L1$\omega^4$ (original) & Constraints~\cite{Dutra2014,Micaela2017} \\
\toprule
 $n_0$ (fm$^{-3}$) & 0.160 & 0.156 &  0.153 &0.150 & 0.155 & 0.164 & 0.148 - 0.170  \\
 $K$ (MeV) &240 &256 & 300 &238 &229 &241 & 220 - 260  \\
$M^{*}_N/M_N$ &0.75 & 0.69 &0.70 & 0.59& 0.61   &0.69 & 0.6 - 0.8    \\ 
$B/A$ (MeV) &16.0 & 16.2 & 16.3 &16.3 & 16.3 & 16.0 &15.8 - 16.5   \\
$S_0$ (Mev) &32.5 &31.7  & 32.5 & 30.5 &31.3  &33.1 & 28.6 - 34.4     \\
 \hline
 $M_N$& 938.93 &938.93 & 938.93& 939.0 & 939.0 & 938.9 & -   \\
 $m_\sigma$&512 &512 & 512 & 497.5 & 492.7 & 512 & - \\
 $m_\omega$&783 &783 & 783 &782.5 & 782.5 &783 & -       \\
 $m_\rho$&770&770 & 770 & 763 & 763 &770 &  - \\
 $g_\sigma$ &8.339 &9.029 & 8.908 &10.136 & 9.670 &8.913 & -    \\
$g_\omega$ & 9.239 &10.597 & 10.609  & 13.020 & 12.316 &10.457 & -  \\
$\kappa$ & 0.00694&0.00414 & 0.00295 & 0.00213& 0.00277 &0.00440 & -   \\
$\lambda$ & -0.00480 &-0,00390 &-0.00107 &-0.00222 &-0.00362 &-0.00480 & - \\
$\xi$ & - & - & -  &0.00133 & 0.00012 & -0.00004 & -   \\
\hline 
\end{tabular}
\caption{Parameter sets used in this work and corresponding saturation properties. } \label{TL1}
\end{center}
\end{table}
\end{center}
\end{widetext}

\begin{center}
\begin{table}[h]
\begin{center}
\begin{tabular}{ccccc}
\toprule
Model &  $L$ (MeV) & $g_\rho$ & $\Lambda_{\omega\rho}$  \\
\toprule
 NL$\rho$ & 44 & 11.037 &  0.0700 \\
NL$\rho$ & 60 & 9.763  &  0.0470 \\
NL$\rho$ & 76& 8.804 &  0.0220        \\
NL$\rho$ &92 &  8.120  &  -0.0010  \\
\hline
L3$\omega\rho$ & 44 & 11.310 &  0.0515 \\
L3$\omega\rho$ & 60 & 9.685  &  0.0344 \\
L3$\omega\rho$ & 76& 8.638 &  0.0171        \\
L3$\omega\rho$ &92 &  7.863  &  0  \\
\hline
GM1 & 44 & 11.536 &  0.0470 \\
GM1 & 60 & 10.086  &  0.0320 \\
GM1 & 76& 9.017 &  0.0165        \\
GM1 &92 &  8.287  &  0.0019  \\
\hline
FSU2H & 44 & 14.049 &  0.0450 \\
FSU2H & 60 & 10.260  &  0.0312 \\
FSU2H & 76& 8.445 &  0.0175        \\
FSU2H &92 &  7.326  &  0.0036  \\
\hline
BigApple & 44 & 12.846 &  0.0440 \\
BigApple & 60 & 10.000  &  0.0310 \\
BigApple & 76& 8.509 &  0.0181        \\
BigApple &92 &  7.567  &  0.0057  \\
\hline
L1$\omega^4$ & 44 & 11.707 &  0.0530 \\
L1$\omega^4$ & 60 & 9.926  &  0.0362 \\
L1$\omega^4$ & 76& 8.818 &  0.0204        \\
L1$\omega^4$ &92 &  5.866  &  0.0050  \\
\toprule
\end{tabular}
\caption{Parameters for four fixed values of the symmetry energy slope within six different parametrizations.} 
\label{T2}
\end{center}
\end{table}
\end{center}

\begin{center}
\begin{table}[ht]
\begin{center}
\begin{tabular}{ccccc}
\toprule
Model &  $L$ (MeV) & $g_\rho$ & $\Lambda_{\omega\rho}$  \\
\toprule
 NL$\rho$ & 79 & 8.647 &  0.0176 \\
L3$\omega\rho$ & 81 & 8.360 &  0.0012 \\
GM1 & 84 & 8.629 &  0.0092 \\
FSU2H &86 &  7.685  &  0.0089  \\

BigApple &86 &  7.919  &  0.0107  \\
L1$\omega^4$ &80 &  8.620  &  0.0165  \\
\toprule
\end{tabular}
\caption{Slope value and the respectively parameters that predict $M_{DU} = 1.35M_\odot$, therefore acting as a model-dependent upper limit for the slope.} 
\label{T3}
\end{center}
\end{table}
\end{center}

Now, the direct URCA (DU) process takes place when the proton fraction exceeds
a critical value $X_{DU}$, which can be evaluated in terms of the leptonic fraction~\cite{Lattimer1991,Dohi2019}:

\begin{equation}
 X_{DU} = \frac{1}{1 + (1 +x_e^{1/3})^3} , \label{xdu}   
\end{equation}
where $x_e = n_e/(n_e + n_\mu)$, and $n_e$, $n_\mu$ are the number densities of the electron and the muon, respectively. Furthermore, as will become clear through the paper, the minimum mass that enables the DU process depends on the slope. In the Tab.~\ref{T3} I display for each parametrization, the slope value that predicts exactly the weak constraint of the DU process, $M_{DU} = 1.35M_\odot$.

\section{Results and discussion}

In order not to saturate the text, I  only present the summarised figures here, which cover all the models and $L$ values. The individualized figures, for each model and each slope value, are presented in the appendix.

\begin{figure}
  \begin{centering}
\begin{tabular}{c}
\includegraphics[width=0.333\textwidth,angle=270]{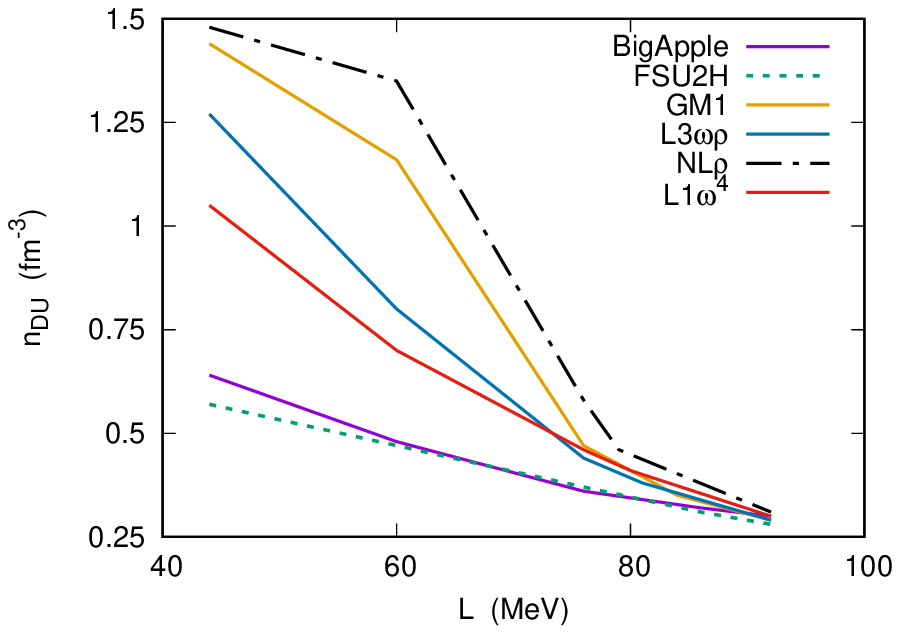} \\
\end{tabular}
\caption{The number density where the proton fraction reaches $X_{DU}$ as a function of the slope (bottom).  (Color online).} \label{Fyp}
\end{centering}
\end{figure}

I begin by calculating the proton fraction ($Y_p$) as a function of the number density for different slope values. The complete results are presented in Fig.~\ref{FA1}, where $X_{DU}$ is defined in Eq~\ref{xdu}. As can be seen, the higher the value of $L$, the higher the proton fraction. One can also notice that for two parameterizations, NL$\rho$ and L3$\omega\rho$, the $Y_p$ has a quick increase, surpassing 0.25 for $L = 92$ MeV. Such behavior, more than a dependence on the physical quantities, seems to be linked to the nature of non-linear $\omega-\rho$ coupling, $\Lambda_{\omega\rho}$. For positive values of $\Lambda_{\omega\rho}$ coupling  (see ref.~\cite{lopescesar} and references therein to additional details), the effective mass of the $\rho$ meson, increases proportional to the number density, reducing its contribution, especially at high densities. For $L = 92$ MeV, the L3$\omega\rho$ has $\Lambda_{\omega\rho}$ equals to zero, while the NL$\rho$ presents a negative value, therefore its effects are reversed.
The density where the proton fraction reaches $X_{DU}$, which I  call by  $n_{DU}$ is presented in Fig.~\ref{Fyp}.

As can be seen, from the qualitative point of view, there is a correlation between the slope and the number density where the proton fraction reaches $X_{DU}$. The lower is the slope, the higher is the $n_{DU}$. From the quantitative point of view, one can see that for low values of the slope, the results present big differences, but for high values of $L$, the results are similar. For instance, within $L = 44$ MeV, $n_{DU}$  reads 1.48 fm$^{-3}$ for NL$\rho$ and  1.44 fm$^{-3}$ for GM1, while assumes $n_{DU} =0.64$ fm$^{-3}$ for BigApple and $n_{DU} =0.57$ fm$^{-3}$ for FSU2H. On the other hand, within $L = 92$ MeV,  $n_{DU} =0.31$, 0.29, 0.30, and 0.28  fm$^{-3}$ for the NL$\rho$, GM1, BigApple and FSU2H respectively. The fact that the results begin to become degenerate at higher values of $L$ was noticed (in a different context) in ref.~\cite{lopes2024PRC}. In their work, the authors show that for higher values of $L$, the critical chemical potential related to hadron-quark phase transition is almost independent of the quark EOS utilized.

\begin{figure}
  \begin{centering}
\begin{tabular}{c}
\includegraphics[width=0.333\textwidth,angle=270]{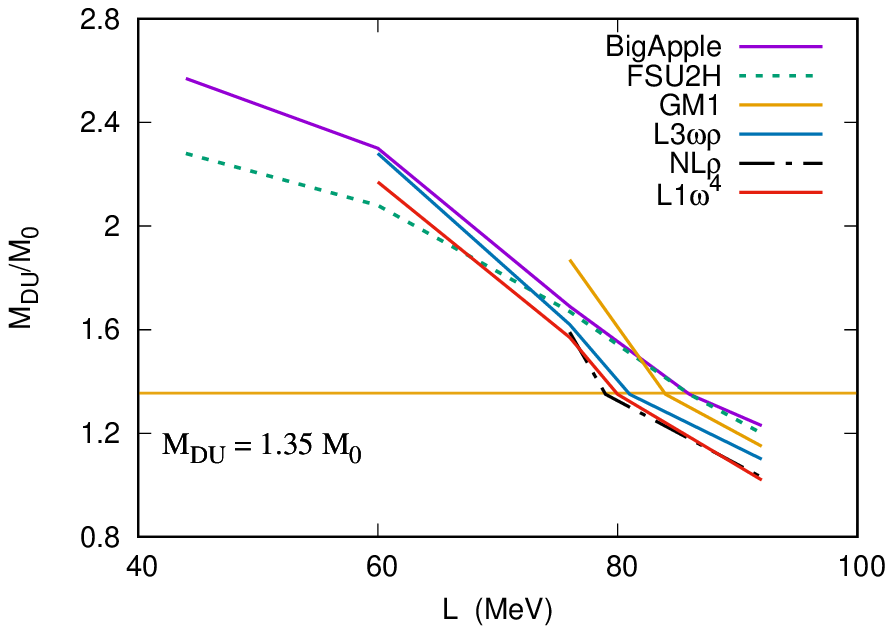} \\
\includegraphics[width=0.333\textwidth,angle=270]{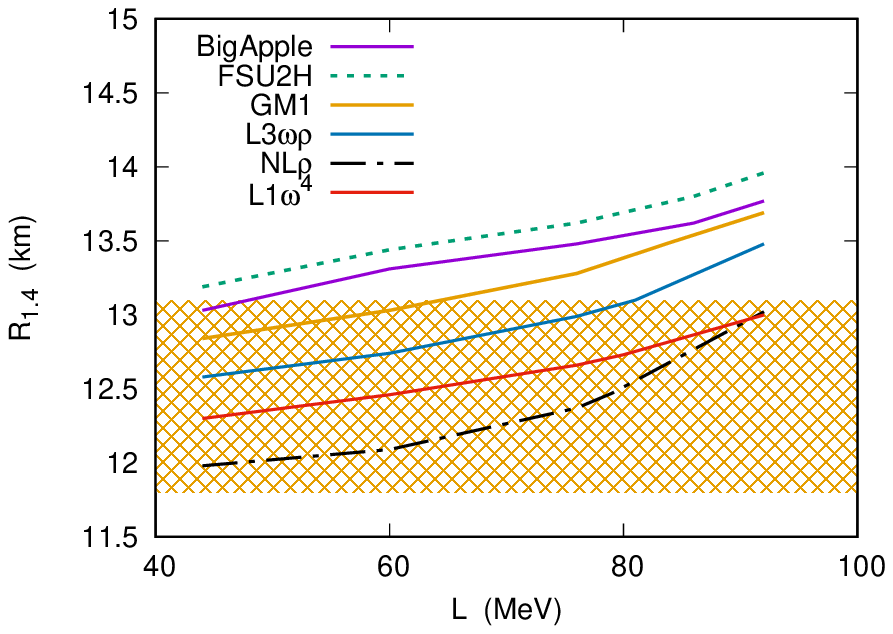} \\
\includegraphics[width=0.333\textwidth,angle=270]{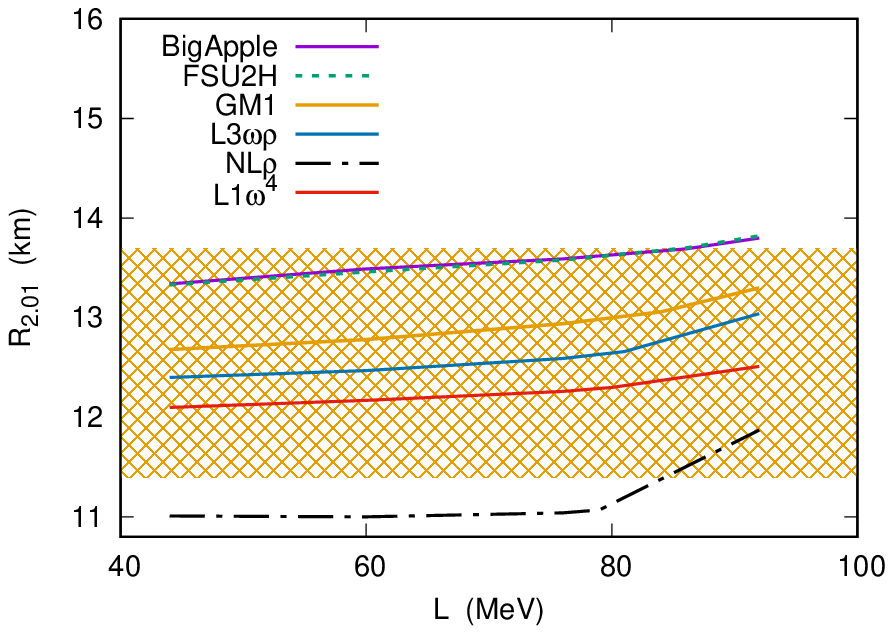} \\
\end{tabular}
\caption{(Top) The minimum mass that enables the DU process as a function of the slope. The horizontal line indicates the weak constraint $M_{DU}~>~1.35M_\odot$. (Middle) The radius of the canonical 1.4$M_\odot$ star as a function of the slope. The hatched area is related to the constraint  $R_{1.4} = 12.45 \pm 0.65$ km~\cite{Miller2021}. (Bottom)  The radius of the 2.01$M_\odot$ star as a function of the slope. The hatched area is the constraint  $R = 12.39^{+1.30}_{-0.98}$ km~\cite{Riley2021}. (Color Online)} \label{Ftov}
\end{centering}
\end{figure}

I now discuss how the slope affects some neutron stars' properties. The complete mass-radius solution is obtained by solving the TOV equations~\cite{TOV}. It is also important to point out that to describe the outer crust and inner crust of the neutron stars, I utilize the Baym-Pethick-Sutherland (BPS) EOS \cite{BPS} and the Baym-Bethe-Pethick (BBP) EOS \cite{BBP}, respectively.   I use the BPS+BBP EoS up to the density of 0.0089 fm$^{-3}$ for all values of $L$, and from this point on, I use the QHD EoS, as suggested in ref.~\cite{Glenbook}. In ref.~\cite{Fortin2016}, the authors compare the BPS+BBP crust EoS with a unified EoS. They show that for the canonical star, there is a variation in the radius of $60 \mathrm{m} <~R_{1.4}~< 150 \mathrm{m}$. For a radius of 13 km, this implies an uncertainty around 1$\%$. This procedure is the same as the one done in ref.~\cite{lopescesar,lopes2024PRC,Lopes2024ApJ}. The complete mass-radius relation is displayed in Fig.~\ref{FA2} in the appendix, while the main results are presented in Fig.~\ref{Ftov}.

At the top of Fig.~\ref{Ftov}, I show the minimum mass that enables the DU process as a function of the slope. That is an important feature because the weak constraint, $M_{DU}~>~ 1.35M_\odot$ must be satisfied. Although the models predict very different stars, the maximum values of the slope are very close to each other, presenting a difference of only 7 MeV, as presented in Tab.~\ref{T3}. Therefore, the results of this table can be used as a model-dependent upper limit of the slope. Moreover, in general, the stiffer the EOS, the higher the upper limit of the slope. However, there is little correlation between the physical quantities at saturation density and $M_{DU}$. One also can see that for $L = 44$ MeV, only the two stiffer EOSs (BigApple and FSU2H) predict the DU process for massive stars. For $L = 60$ MeV, L3$\omega\rho$ and L1$\omega^4$ also predict the DU process for massive stars. Finally, for $L~\geq$ 76 MeV, all models predict the DU process at least for massive stars. 

In the middle of the Fig.~\ref{Ftov}, I display the radius of the canonical star altogether with the constraint $R_{1.4} = 12.45 \pm 0.65$ km, as presented in ref.~\cite{Miller2021}. There is an increase in the radius with the slope for all parametrizations. The FSU2H does not satisfy this constraint for either value of $L$. BigApple satisfies only for $L = 44$ MeV. On the other hand, NL$\rho$ and L1$\omega^4$ satisfy this constraint for all values of $L$. If, however, one uses a more conservative constraint, as $R_{1.4}~<$ 13.6 km as suggested in ref.~\cite{Annala2018PRL}, then FSU2H can be satisfied up to $L = 60$ MeV. In the same sense, BigApple can be satisfied up to 76 MeV, while the L3$\omega\rho$ is satisfied for all values of $L$. Here again, I do not see any correlation between the radius of the canonical star and the physical quantities at saturation point, except the well-known fact that increasing the slope also increases the radius. There is, nevertheless, a correlation between a stiffer EOS and the radius of the canonical star. The exception is the FSU2H, which is softer than the BigApple but has a higher radius for the canonical star with the same value of $L$. But this can be explained by the coupling constant with the $\omega$ meson, $g_{\omega}$, and the strength of its self-interaction, $\xi$. While increasing $g_\omega$ stiffens the EOS, increasing $\xi$ softens the EOS at high densities. The FSU2H has both: the largest value of $g_\omega$ and $\xi$, indicating a stiffened EOS at moderate densities, but it begins to soften at higher densities. Similarly, the negative value of $\xi$ for the L1$\omega^4$ parametrization produces a soft EOS at moderate density but a stiffer one at higher densities. 

At the bottom of Fig.~\ref{Ftov}, I plot the radius of the 2.01$M_\odot$ star,  which is not only the most probable mass value of PSR J0348+0432~\cite{Antoniadis} but also the lower limit of PSR J0740+6620, whose gravitational mass is 2.08 $\pm$ 0.07 $M_\odot$~\cite{Miller2021,Riley2021}. Therefore, any equation of state (EOS) unable to reach at least 2.01 $M_\odot$ must be ruled out. Altogether, there is also the constraint to the radius of the  PSR J0348+0432, $R = 12.39^{+1.30}_{-0.98}$ km. One can see that the correlation between the slope and the radius of the 2.01$M_\odot$ star is much weaker than the correlation for the canonical star. It also can be seen that most parametrizations agree with the constraint $R = 12.39^{+1.30}_{-0.98}$. The exception is the NL$\rho$, which produces too low values. Although it can be satisfied for $L = 92$ MeV, the upper limit for this parametrization based on $M_{DU}~>~1.35M_\odot$ is 79 MeV.

\begin{figure}[b]
  \begin{centering}
\begin{tabular}{c}
\includegraphics[width=0.333\textwidth,,angle=270]{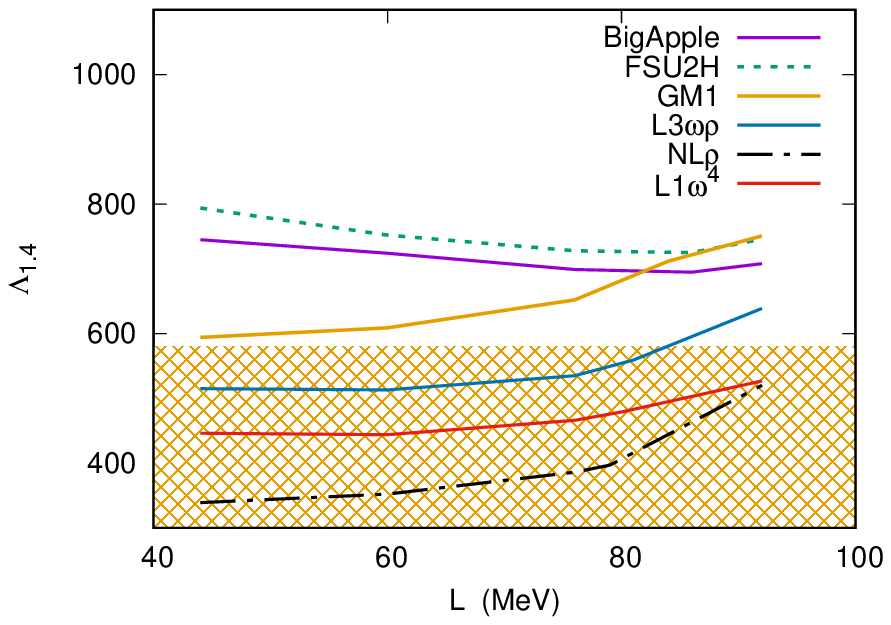} \\
\end{tabular}
\caption{ The relation between the dimensionless tidal parameter for the canonical star, $\Lambda_{1.4}$, and the slope (Color online). } \label{Ftid}
\end{centering}
\end{figure}

 I now discuss a recent and fierce constraint presented in ref.~\cite{Dex2024arxiv}. The authors indicate that during the April APS meeting, the NICER group reported that the   PSR J0437-4715, with a mass of $M_{J0437}$ =  1.418$M_\odot$,  has a radius of only $11.36^{+0.95}_{-0.63}$ km. In this case, only the two lower values of $L$ for the NL$\rho$ and only $L = 44$ MeV for the L1$\omega^4$ can satisfy such a strong constraint. However, as mentioned, the NL$\rho$ fails to fulfill the radius of the PSR J0740+6620, making the L1$\omega^4$ within $L = 44$ MeV the sole parametrization able to satisfy both constraints. Moreover, with a maximum mass of 2.29$M_\odot$ even the speculative black widow pulsars PSR J0952-0607 with a mass  $M = 2.35\pm0.17 M_\odot$~\cite{romani} can be reached.

To finish,  I discuss the influence of the slope in the dimensionless tidal parameter $\Lambda$. The tidal deformability of a compact object is a single parameter that quantifies how easily the object is deformed when subjected to an external tidal field. A larger tidal deformability indicates that the object is easily deformable. On the opposite side, a compact object with a smaller tidal deformability parameter is smaller,  more compact, and more difficult to deform. It is defined as:

\begin{equation}
\Lambda = \frac{2 k_2}{3C^5} , \label{etid}
\end{equation}
where $C = M/R$ is the neutron star compactness. The parameter $k_2$ is called the Love number and is related to the metric perturbation. A complete discussion about the Love number and its calculation is both, very extensive and well-documented in the literature. Therefore, it is out of the scope of the present work. I refer the interested reader to see ref.~\cite{Eanna2008,Hinderer_2008,Chat2020,Flores2020} and the references therein.

In Fig.~\ref{Ftid}, I show the dimensionless tidal parameter for the canonical star, $\Lambda_{1.4}$, as a function of the slope, altogether with the constraint 70 $<~\Lambda_{1.4}~<$  580 coming from the GW170817 event detected by the VIRGO/LIGO gravitational wave observatories~\cite{AbbottPRL}. The complete results for $\Lambda$ are presented in Fig.~\ref{FA3} in the appendix. One can see that there are two different behaviors. Within the NL$\rho$ and the GM1, the $\Lambda_{1.4}$ monotonically increases with the slope. The other four parametrizations present a minimum for $\Lambda_{1.4}$. For the L3$\omega\rho$ and L1$\omega^4$ this minimum happens at $L = 60$ MeV, while for the FSU2H and BigApple, it happens at $L = 86$ MeV. The fact that some parametrization presents a decreasing $\Lambda_{1.4}$ with the slope is related to the competition between the compactness $C$ and the Love number $k_2$, as pointed out in ref.~\cite{lopescesar}. Although increasing the slope also increases the radius and consequently reduces the compactness, it also reduces the Love number $k_2$. The lower values of $L$, actually present large values of $k_2$. I display the Love number $k_2$ for different values of $L$ in Fig.~\ref{FA4} in the appendix. The large value of $k_2$ within $L = 44$ MeV for both, BigApple and FSU2H, explains the large value of $\Lambda_{1.4}$ even for low values of the slope. 

As in the case of $R_{1.4}$, the $\Lambda_{1.4}$ seems to be more linked to the stiffness of the EOSs than to the physical quantities at saturation density. Nevertheless, the compressibility $(K)$ may play a secondary role in the tidal parameter. For instance, GM1 within $L = 60$ MeV and the L3$\omega\rho$ within $L = 76$ MeV present similar maximum masses as well as very similar radii for the canonical star. Still, the GM1 predicts $\Lambda_{1.4}$ significantly larger than the L3$\omega\rho$. Another parametrization close to the previous two is the L1$\omega^4$ within $L = 92$ MeV, which corroborates the possible secondary role of $K$.

Concerning the constraint 70 $<~\Lambda_{1.4}~<$  580, it can be seen that the three stiffer EOSs fail to satisfy it for all values of $L$. On the other hand, the two softer ones satisfy it for all values of $L$. The L3$\omega\rho$ fulfills this constraint up to 81 MeV, which coincides with the upper limit due to $M_{DU}~>1.35M_\odot$. A complete table, with the values of all quantities discussed in the text, as well as some constraints, is presented as Tab.~\ref{T4}.

\begin{widetext}
\begin{center}
\begin{table}[hb]
\begin{center}
\scriptsize
\begin{tabular}{cc|cccccc|cccccc}
\toprule
Model &  $L$ (MeV) & $n_{DU}$ (fm$^{-3}$) & $M_{DU}/M_\odot$ & $R_{1.4}$~ & $R_{2.01}$ ~& $\Lambda_{1.4}$ & $M_{max}/M_\odot$ & $M_{DU}>1.35$ & NICER$_{1.4}$ & $R_{1.4}<13.6$ & J0740+ & GW170817 & J0437-\\
\toprule
 NL$\rho$ & 44 & 1.48 &  - & 11.98 & 11.01 & 339 & 2.05 & YES &  YES & YES & NO & YES & YES \\
NL$\rho$ & 60 & 1.35  & - & 12.09 & 11.00 & 352 & 2.05 & YES & YES & YES & NO & YES & YES \\
NL$\rho$ & 76& 0.58 &  1.59& 12.37& 11.04 & 386 & 2.04 & YES & YES & YES &NO & YES & NO        \\
NL$\rho$ & 79& 0.46 & 1.35 & 12.47 & 11.07 & 397 & 2.04 & YES & YES & YES & NO & YES & NO     \\
NL$\rho$ &92 &  0.31  & 1.03 & 13.02 & 11.87 & 520 & 2.11 & NO & YES & YES & YES & YES & NO  \\
\hline
L3$\omega\rho$ & 44 & 1.27 &  -  & 12.58 & 12.40 & 515 & 2.31 & YES & YES & YES & YES & YES &  NO \\
L3$\omega\rho$ & 60 & 0.80  & 2.28 & 12.74 & 12.47 & 513 & 2.30 & YES & YES & YES & YES & YES &  NO \\
L3$\omega\rho$ & 76& 0.44 &  1.62 & 12.99 & 12.59 & 535 & 2.30 & YES & YES & YES & YES & YES & NO       \\
L3$\omega\rho$ & 81& 0.38&  1.35 & 13.10 & 12.66 & 559 & 2.30 & YES & YES & YES & YES & YES  & NO        \\
L3$\omega\rho$ &92 &  0.29  & 1.10 & 13.48 & 13.04 & 639 & 2.34 & NO & NO & YES & YES & NO  &  NO  \\
\hline
GM1 & 44 & 1.44 &  - & 12.84 & 12.68 & 594 & 2.33 & YES & YES & YES & YES & NO & NO  \\
GM1 & 60 & 1.16  & - & 13.03 & 12.78 & 609 & 2.32 & YES & YES & YES & YES & NO & NO \\
GM1 & 76&  0.47 &  1.87 & 13.28 & 12.94 & 652 & 2.32 & YES & NO & YES & YES & NO & NO         \\
GM1 & 84& 0.35 &  1.35  & 13.49 & 13.06 & 712 & 2.32 & YES & NO & YES & YES & NO & NO      \\
GM1 &92 &  0.29  &  1.15 & 13.69 & 13.30 & 751 & 2.34 & NO & NO & NO & YES & NO & NO   \\
\hline
FSU2H & 44 & 0.57 & 2.28 & 13.19 & 13.33 & 794 & 2.38 & YES & NO & YES & YES & NO & NO  \\
FSU2H & 60 & 0.47  & 2.08 &  13.44 & 13.46 & 752 & 2.38 & YES & NO & YES & YES & NO & NO \\
FSU2H & 76& 0.36 &  1.67 & 13.62 & 13.58 & 728 & 2.38 & YES & NO & NO & YES & NO & NO        \\
FSU2H & 86& 0.31 &  1.35 & 13.80 & 13.70 & 725 & 2.39 & YES & NO & NO & YES & NO & NO        \\
FSU2H &92 &  0.28  & 1.20 & 13.96 & 13.82 & 746 & 2.40 & NO & NO & NO & NO & NO & NO  \\
\hline
BigApple & 44 & 0.64 &  2.57 & 13.03 & 13.34 & 745 & 2.60 & YES & YES & YES & YES & NO & NO\\
BigApple & 60 & 0.48  &  2.30 & 13.31 & 13.49  & 724 & 2.60 & YES & NO & YES & YES & NO & NO \\
BigApple & 76& 0.36 & 1.69 & 13.48 &  13.59 & 699 & 2.60 & YES & NO & YES & YES & NO & NO    \\
BigApple & 86& 0.32 &  1.35 & 13.62 & 13.59 & 695 &2.60 & YES & NO & NO & YES & NO & NO      \\
BigApple &92 &  0.30 & 1.23  &  13.77 & 13.80 & 708 &  2.60 & NO & NO & NO & NO & NO & NO  \\
\hline
L1$\omega^4$ & 44 & 1.05 & - &  12.30 & 12.10 & 446 & 2.29 & YES & YES & YES & YES & YES & YES \\
L1$\omega^4$ & 60 & 0.70  & 2.17 & 12.46 & 12.17 & 444 & 2.28 & YES & YES & YES & YES & YES &   NO  \\
L1$\omega^4$ & 76&  0.46 &  1.57 & 12.66  & 12.26 & 466 &  2.28 & YES & YES & YES & YES & YES  & NO    \\
L1$\omega^4$ & 80& 0.41 & 1.35 &  12.73 & 12.30 & 479 &   2.28 & YES & YES & YES & YES & YES & NO      \\
L1$\omega^4$ &92 &  0.30 & 1.02 & 13.00 & 12.51  &  527 & 2.28 & NO & YES & YES & YES & YES &  NO  \\
\toprule 
\end{tabular}
\caption{ Some of the neutron stars' main properties and constraints. The radii are presented in km.} 
\label{T4}
\end{center}
\end{table}
\end{center}
\end{widetext}

\section{Conclusions}

Using six different parameterizations of the QHD, I study the influence of the slope on neutron star properties and explore the model dependency of the results. The main conclusions are:

\begin{itemize}

\item The proton fraction, $Y_p$,  is dependent on the slope. The higher the slope, the lower is the proton fraction at high densities. The qualitative results are the same for all parameterizations.

    \item The non-linear coupling of the $\omega-\rho$ mesons, $\Lambda_{\omega\rho}$ seems to play a more relevant role in the $Y_p$ than the physical quantities at saturation density. If its value is turned to zero or even becomes negative, the $Y_p$ presents a quick increase.

    \item The minimum mass that enables DU process ($M_{DU}$) decreases with the slope. If one imposes the weak constraint $M_{DU}~\leq~1.35M_\odot$, a model-dependent upper limit for the slope appears. Although it is model-dependent, the variation of the upper limit of $L$ is very low, only 7 MeV.

    \item The radius of the canonical 1.4$M_\odot$ star always increases with the slope. The same is true for the radius of the 2.01$M_\odot$ star, although it is less sensible. The absolute values of the $R_{1.4}$ seem to be more related to the strength of the model than to the physical quantities at saturation density. 

    \item The revised constraint presented by the NICER group, $R_{1.4} = 12.45 \pm 0.65$ km~\cite{Miller2021}, cannot be satisfied for the FSU2H neither the BigApple for any value of $L$. On the other hand, the constraint related to the  PSR J0740+6620, whose gravitational mass is 2.08 $\pm$ 0.07 $M_\odot$ and radius $R = 12.39^{+1.30}_{-0.98}$ km~\cite{Miller2021,Riley2021} cannot be satisfied by the NL$\rho$.

    \item The L1$\omega^4$ within $L = 44$ MeV is the only parameterization in this work that is able to simultaneously satisfy the strong constraint reported in ref.~\cite{Dex2024arxiv} related to the PSR J0437-4715 ($M_{J0437}$ =  1.418$M_\odot$,   $R~=~11.36^{+0.95}_{-0.63}$ km), altogether with the PSR J0740+6620. Indeed, even the speculative black widow pulsars PSR J0952-0607 with a mass  $M = 2.35\pm0.17 M_\odot$~\cite{romani} can be explained.

    \item The dimensionless tidal parameter of the canonical star can increase or decrease with the slope, depending on the model used. The stiffer EOSs (GM1, FSU2H, and BigApple) fail to reproduce the constraints coming from the LIGO/VIRGO gravitational wave observatories, $70~<~\Lambda_{1.4}~<580$ for all values of $L$. On the other hand, this constraint is satisfied for all values of $L$ for the softer EOSs (NL$\rho$ and L1$\omega^4$). Moreover, the compressibility, $K$, seems to play a secondary role in this quantity.

    \item Ultimately, very large values of the slope, as suggested by the PREX group~\cite{Tagami2022}, seem to be very unlike. Indeed, values of $L$ around 90 MeV are enough to produce $M_{DU}~<1.35M_\odot$. Moreover, for most of the parametrization, they are also beyond the upper limit of the radius and tidal deformability of the canonical star. The results presented here are a little stronger than those presented in ref.~\cite{lopes2024PRC}, related to the hadron-quark phase transition, which rules out $L~\geq$ 100.

\end{itemize}

$$ $$

\textbf{Acknowledgements:} L.L.L. was partially supported by CNPq Universal Grant No. 409029/2021-1. 

\appendix

\counterwithin{figure}{section}
\section{Complementary Figures}

In order to not saturate the main text, the individualized results for each parametrization and for each value of $L$ is presented here in the appendix.

\begin{figure*}[hb]
\begin{tabular}{ccc}
\centering 
\includegraphics[scale=.58, angle=270]{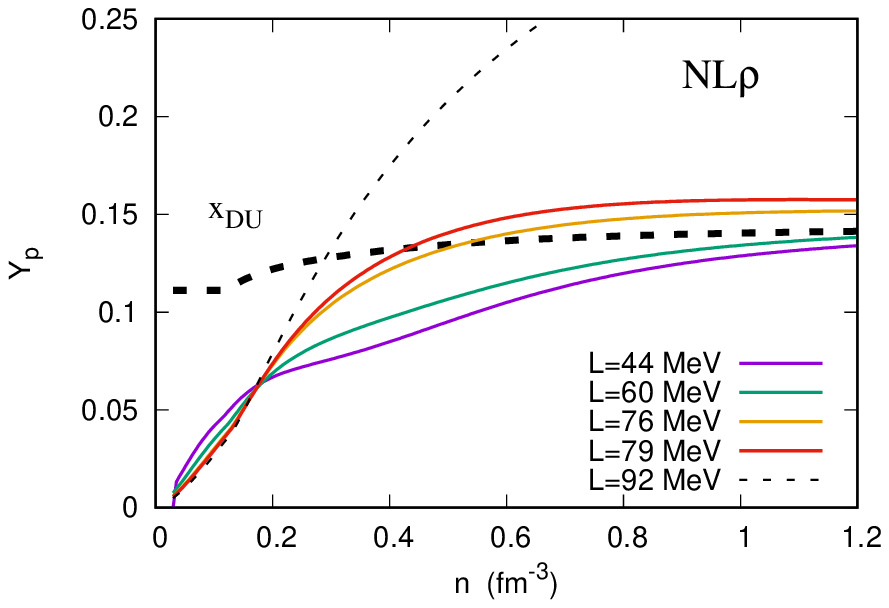} &
\includegraphics[scale=.58, angle=270]{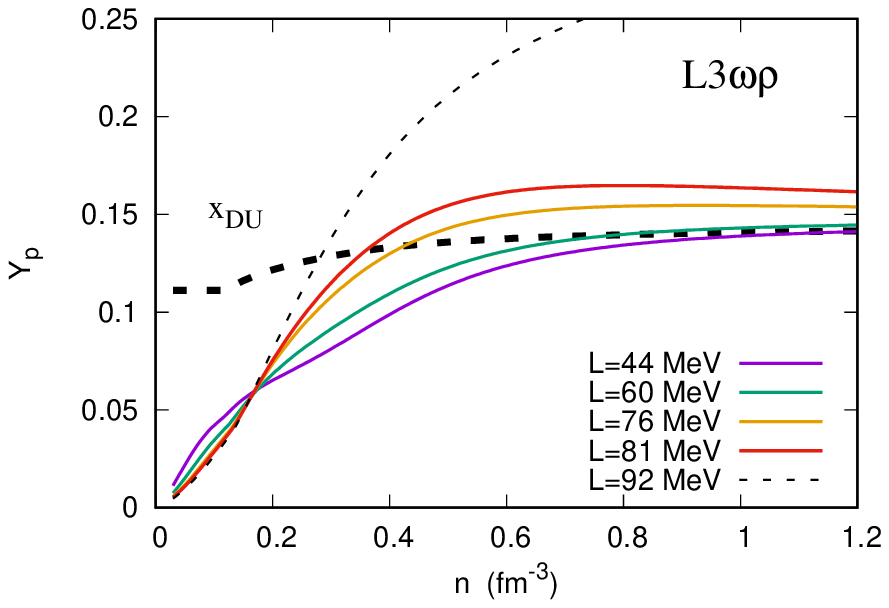} \\
\includegraphics[scale=.58, angle=270]{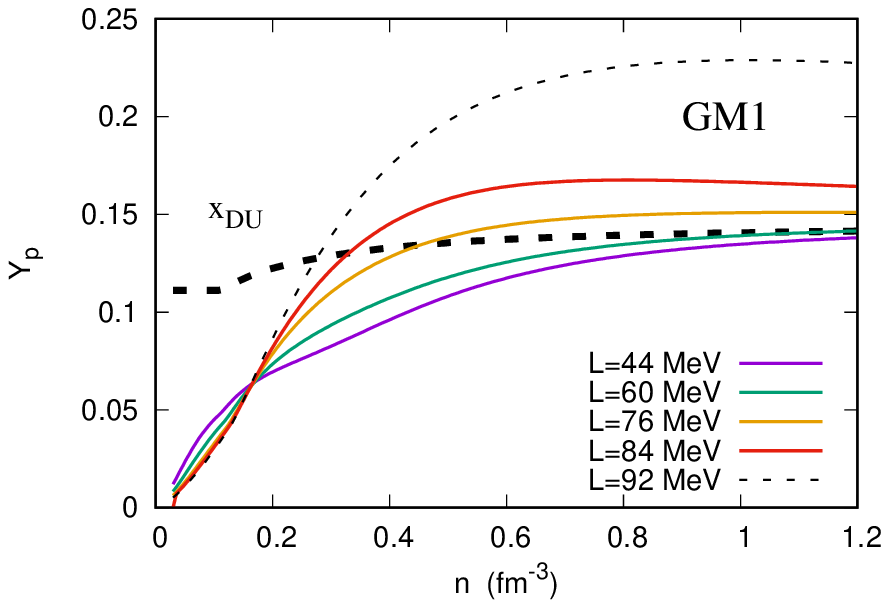} &
\includegraphics[scale=.58, angle=270]{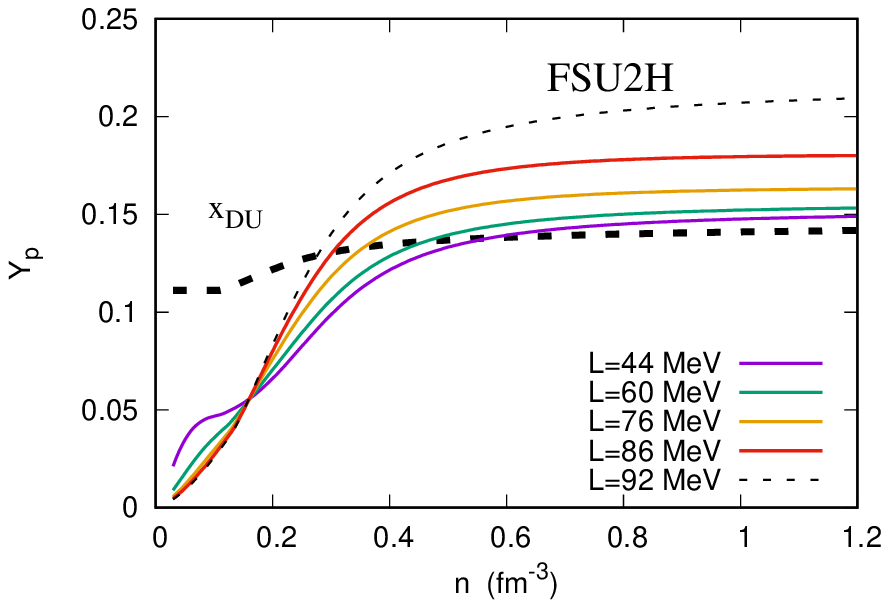}\\
\includegraphics[scale=.58, angle=270]{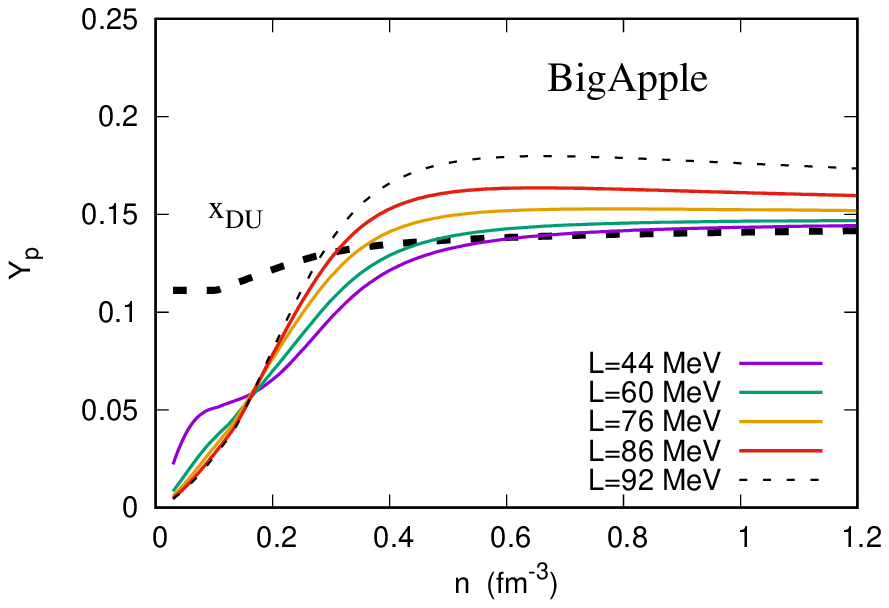} &
\includegraphics[scale=.58, angle=270]{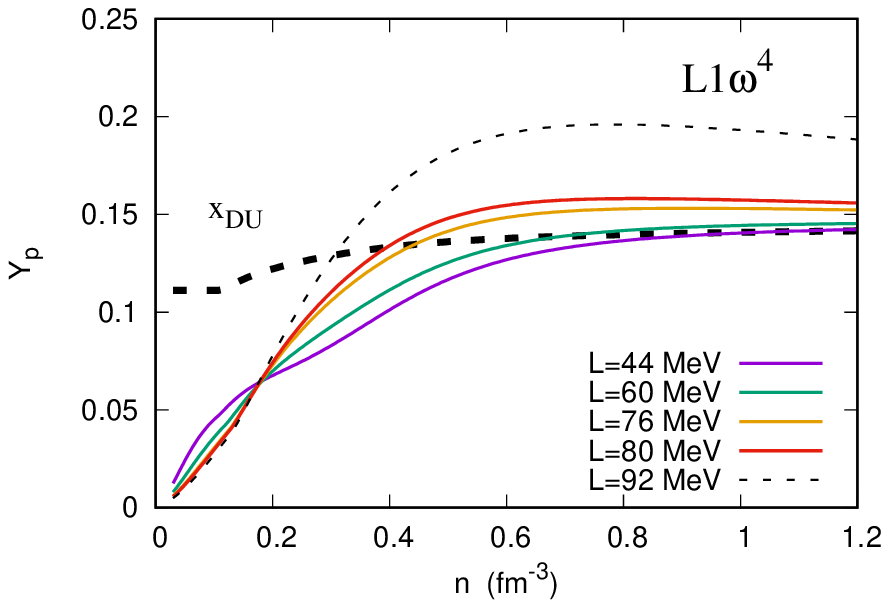}\\
\end{tabular}
\caption{Proton fraction as a function of the number density
for different values of $L$. } \label{FA1}
\end{figure*}

\begin{figure*}[t]
\begin{tabular}{ccc}
\centering 
\includegraphics[scale=.58, angle=270]{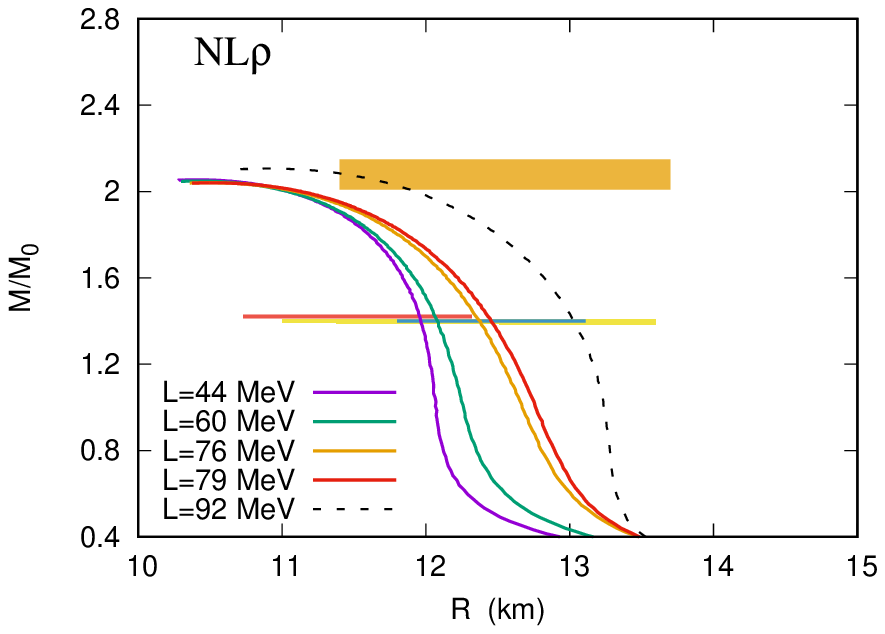} &
\includegraphics[scale=.58, angle=270]{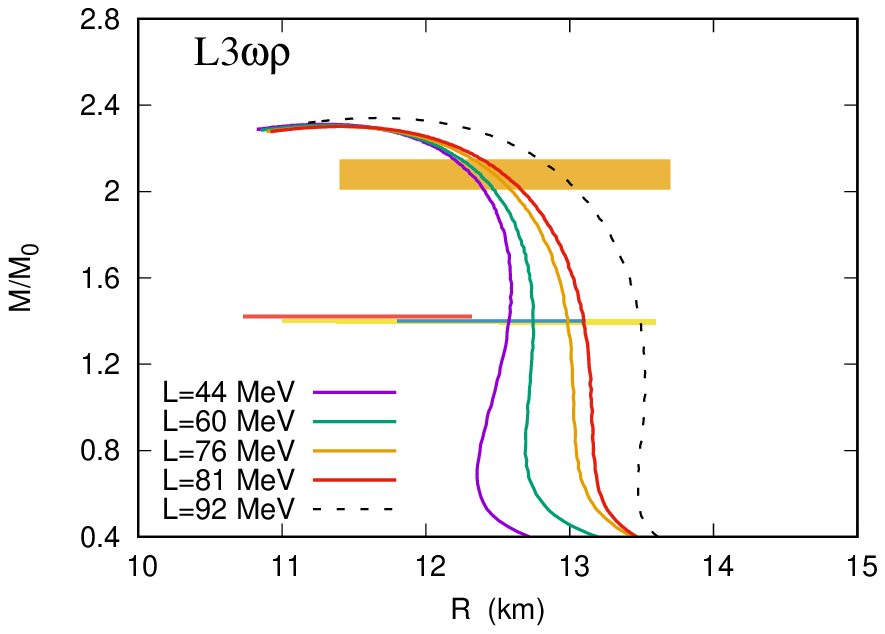} \\
\includegraphics[scale=.58, angle=270]{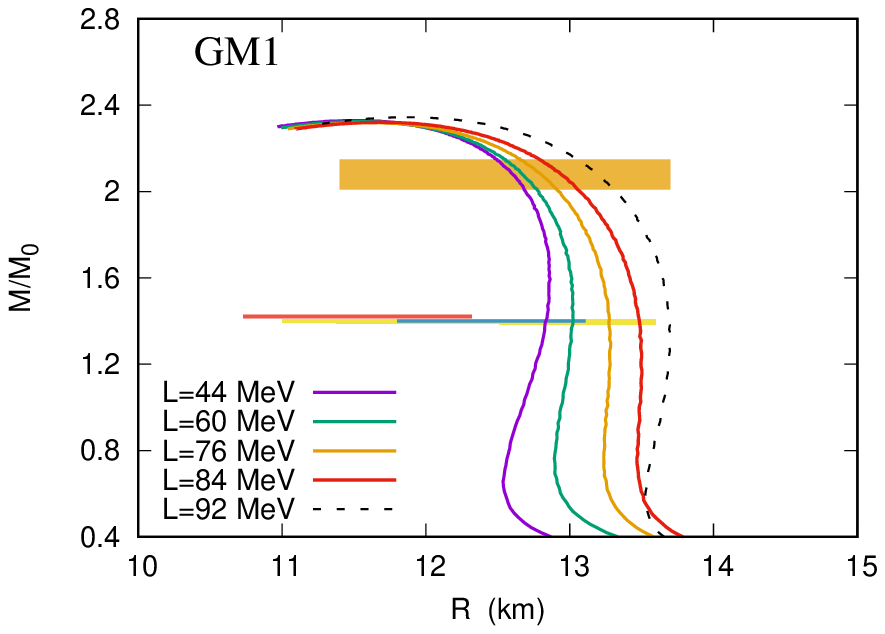} &
\includegraphics[scale=.58, angle=270]{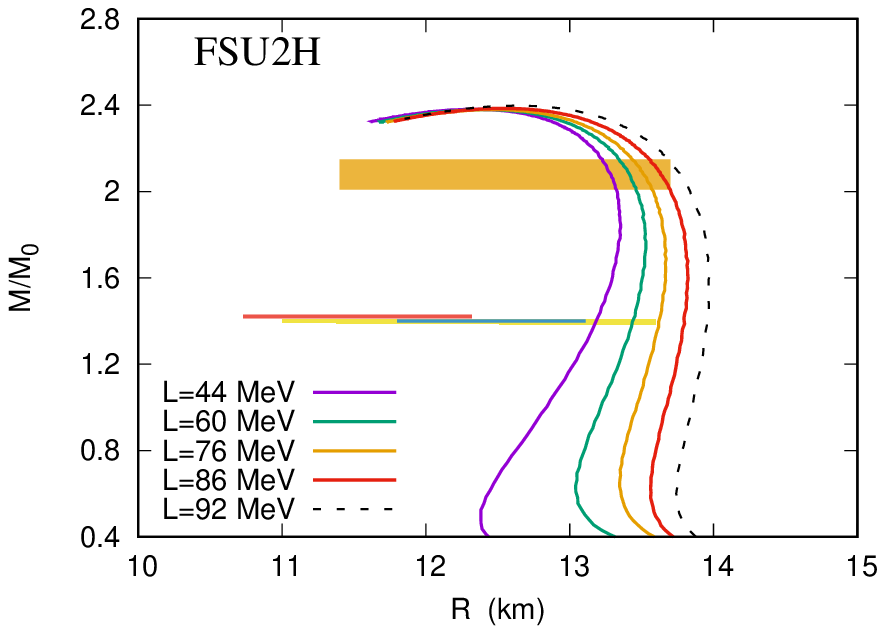}\\
\includegraphics[scale=.58, angle=270]{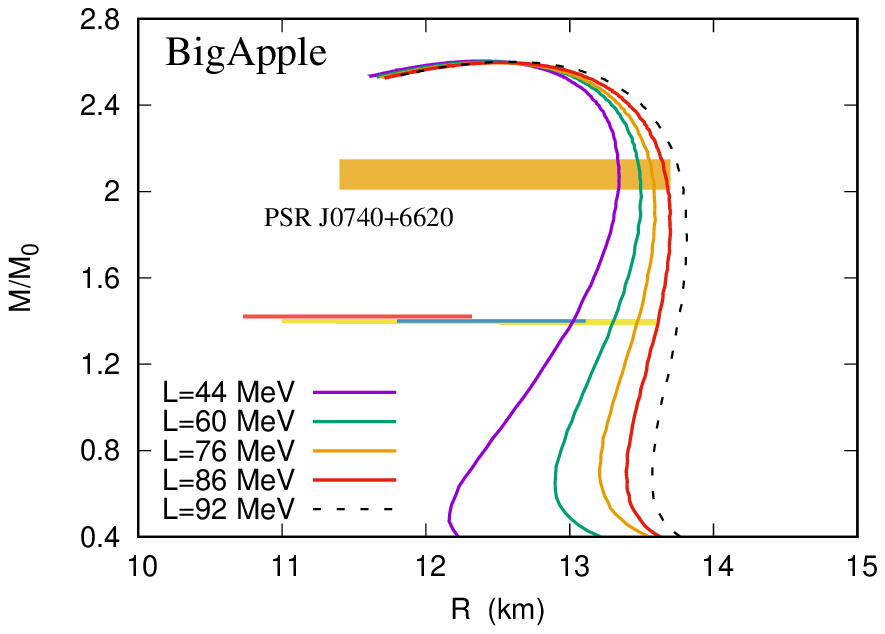} &
\includegraphics[scale=.58, angle=270]{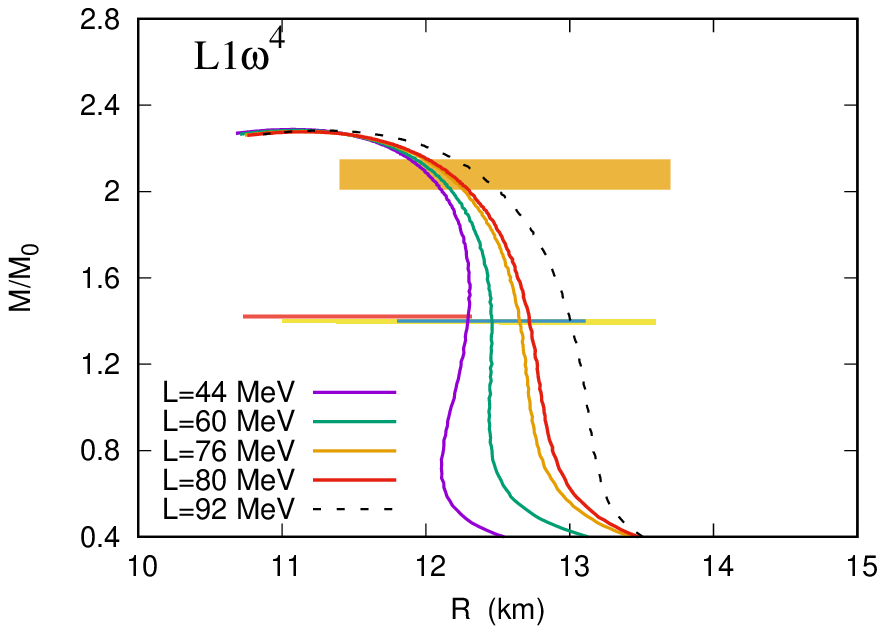}\\
\end{tabular}
\caption{Mass-radius relation for different values of $L$. The hatched area are  constraints related to the PSR J0740+6620, with M = $2.08 \pm 0.07 M_\odot$ and and $R = 12.39^{+1.30}_{-0.98}$ km (dark yellow)~\cite{Riley2021}; PSR J0437-4715 with $M = 1.418M_\odot$ and $R = 11.36^{+0.95}_{-0.63}$ km (red)~\cite{Dex2024arxiv}; $R_{1.4} = 12.45 \pm 0.65$ km (blue)~\cite{Miller2021} and $R_{1.4}~<$ 13.6 km (light yellow)~\cite{Annala2018PRL}.} \label{FA2}
\end{figure*}

\begin{figure*}[t]
\begin{tabular}{ccc}
\centering 
\includegraphics[scale=.58, angle=270]{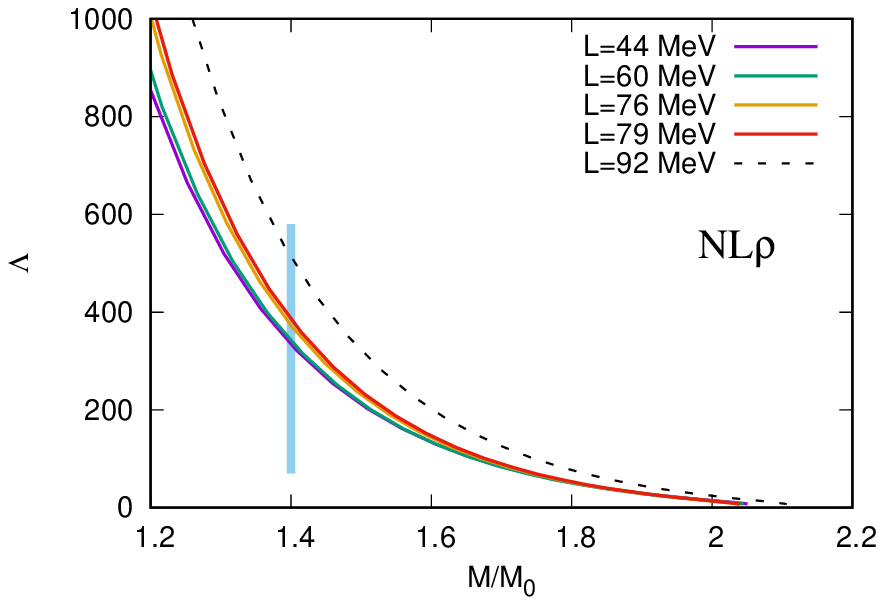} &
\includegraphics[scale=.58, angle=270]{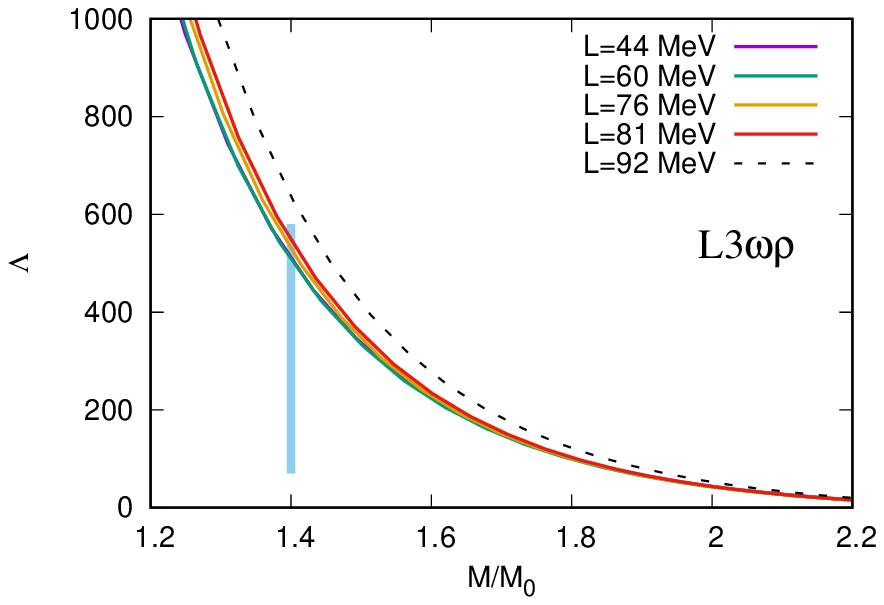} \\
\includegraphics[scale=.58, angle=270]{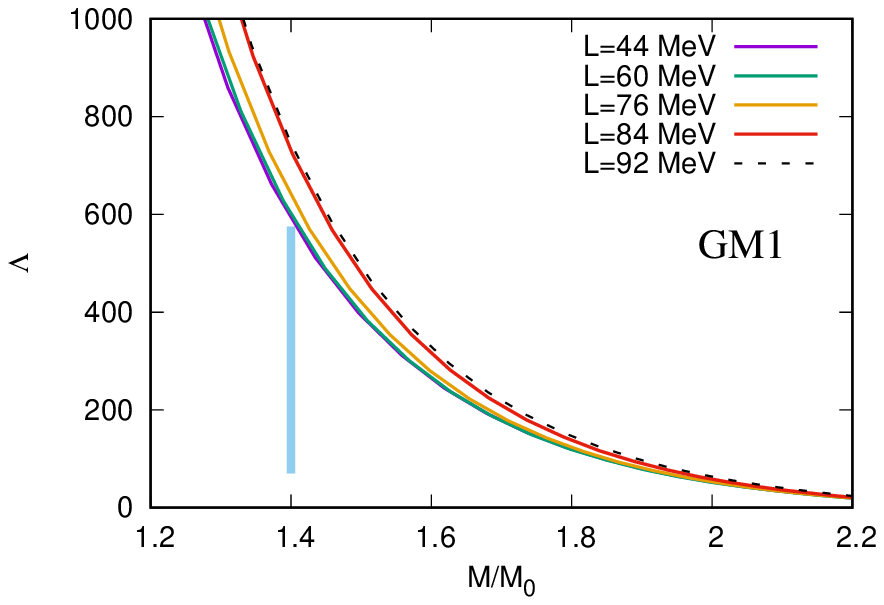} &
\includegraphics[scale=.58, angle=270]{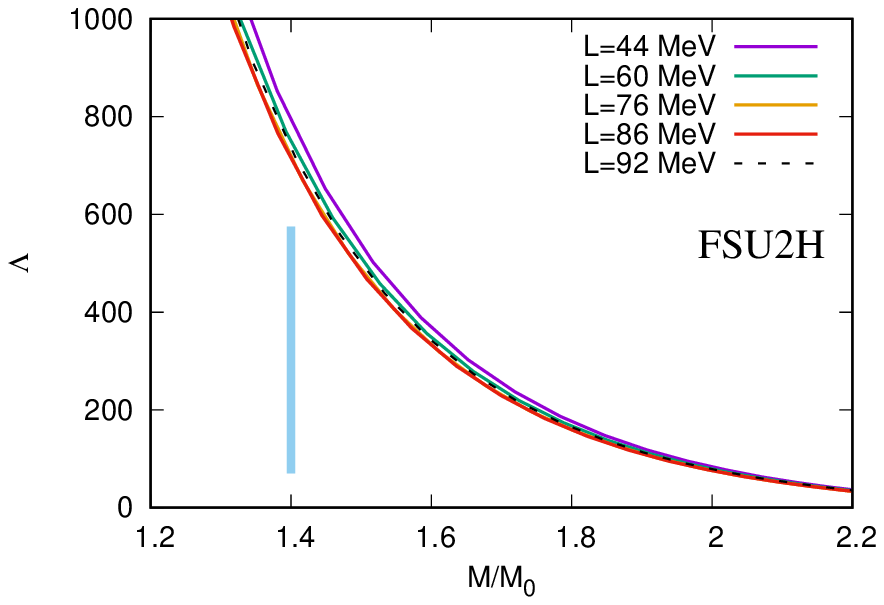}\\
\includegraphics[scale=.58, angle=270]{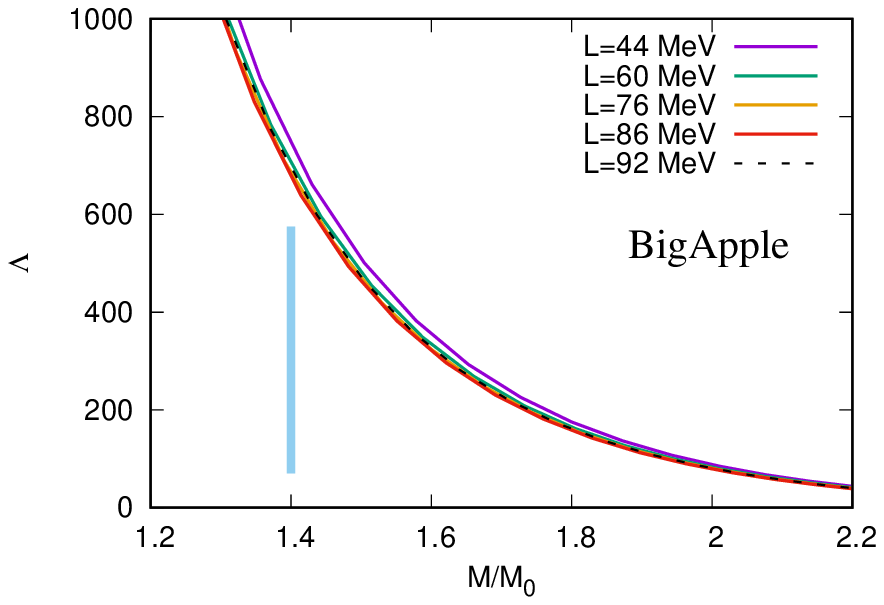} &
\includegraphics[scale=.58, angle=270]{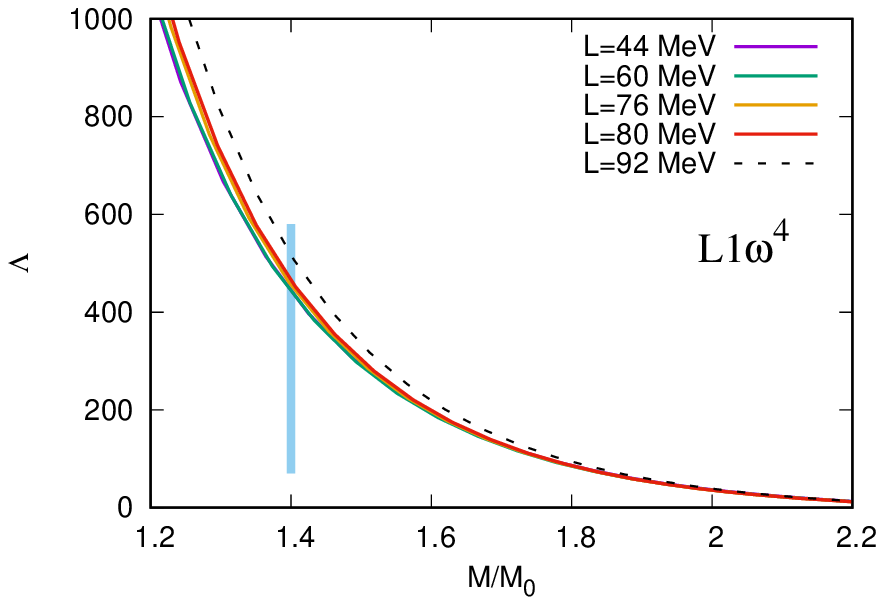}\\
\end{tabular}
\caption{Dimensionless tidal parameter $\Lambda$ for different values of $L$. The hatched area is related to the GW170817 event, 70 $<~\Lambda_{1.4}~<$ 580~\cite{AbbottPRL}.} \label{FA3}
\end{figure*}

\begin{figure*}[t]
\begin{tabular}{ccc}
\centering 
\includegraphics[scale=.58, angle=270]{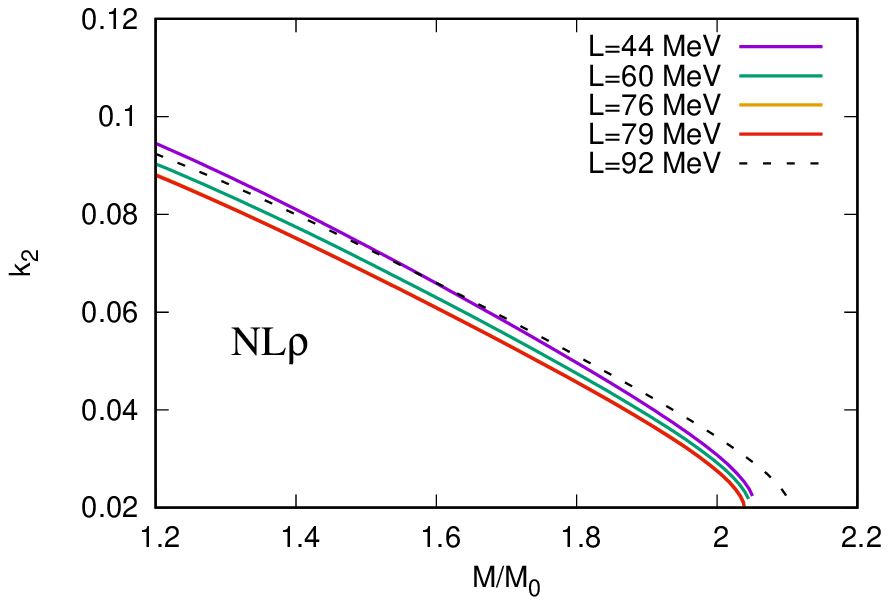} &
\includegraphics[scale=.58, angle=270]{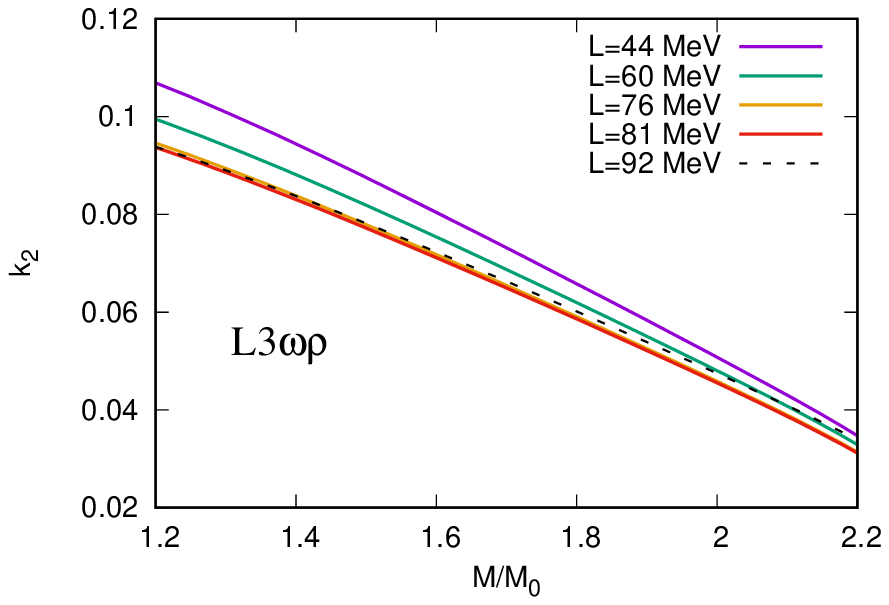} \\
\includegraphics[scale=.58, angle=270]{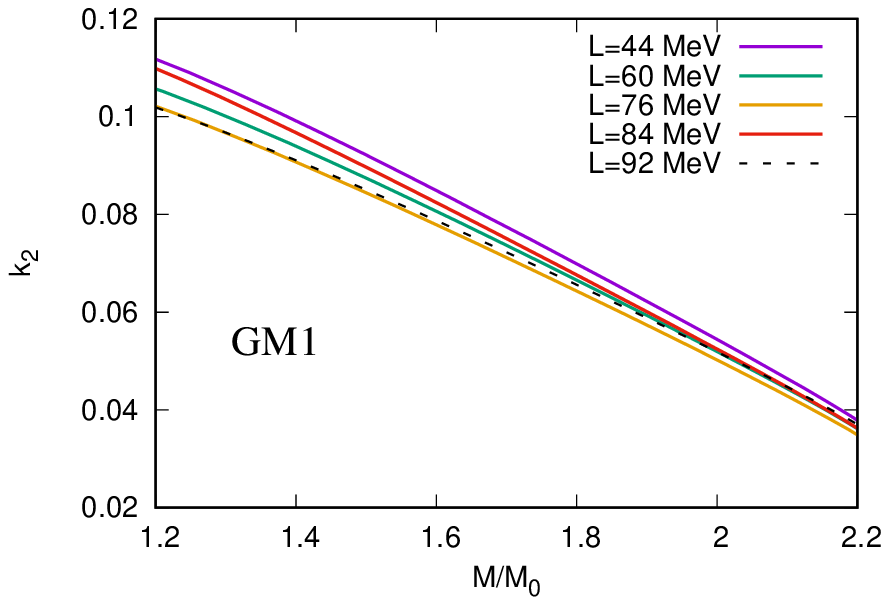} &
\includegraphics[scale=.58, angle=270]{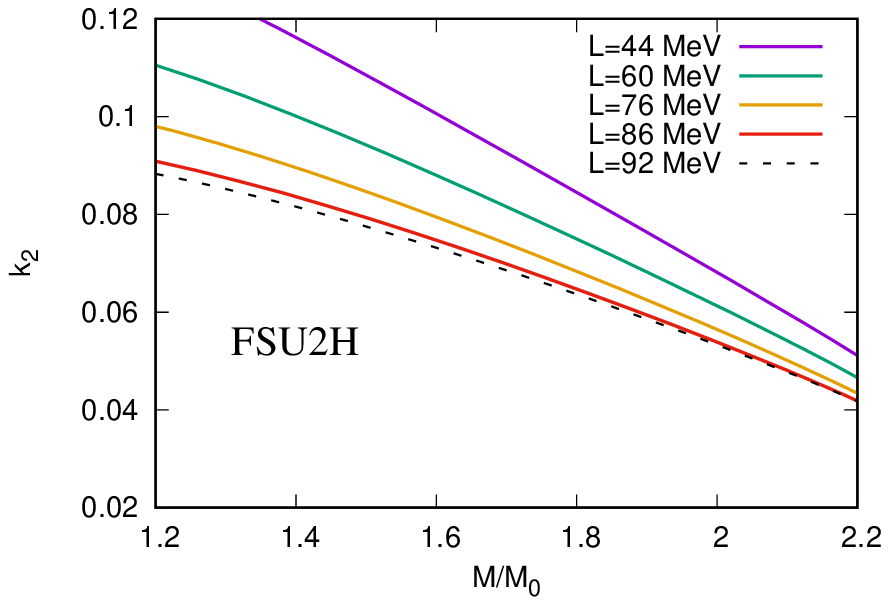}\\
\includegraphics[scale=.58, angle=270]{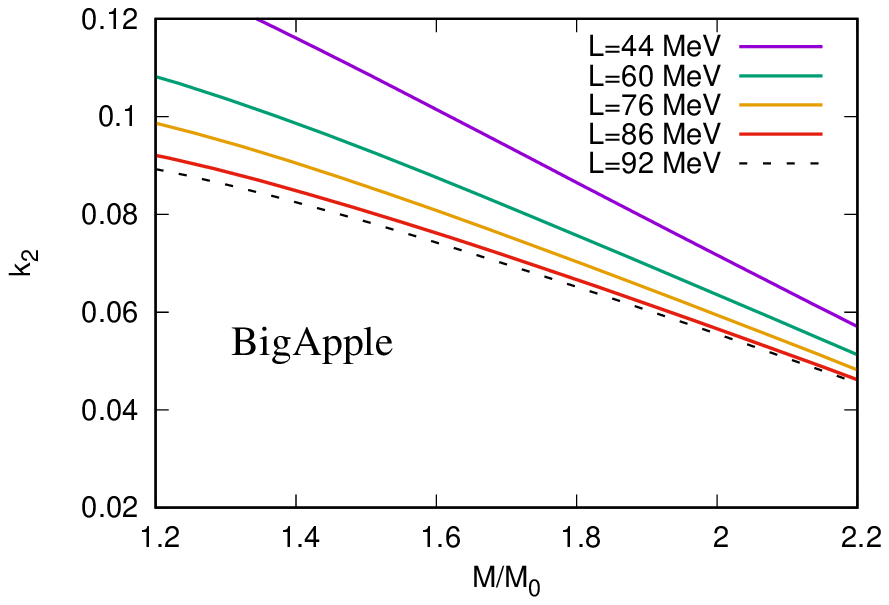} &
\includegraphics[scale=.58, angle=270]{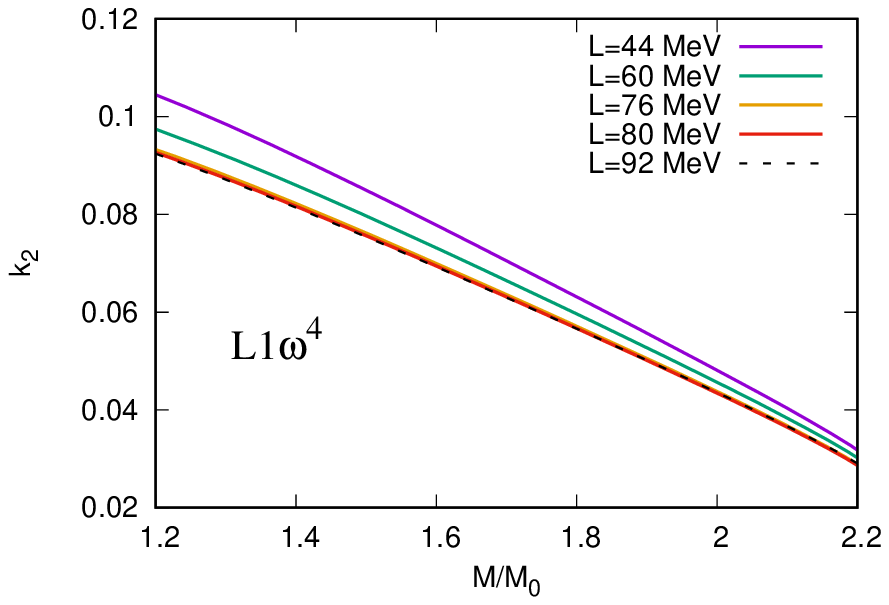}\\
\end{tabular}
\caption{Love number $k_2$ for different values of $L$. In general, the lower the slope, the higher the $k_2$.} \label{FA4}
\end{figure*}

$$ $$

\clearpage

\bibliography{aref}

\begin{thebibliography}{53}%
\makeatletter
\providecommand \@ifxundefined [1]{%
 \@ifx{#1\undefined}
}%
\providecommand \@ifnum [1]{%
 \ifnum #1\expandafter \@firstoftwo
 \else \expandafter \@secondoftwo
 \fi
}%
\providecommand \@ifx [1]{%
 \ifx #1\expandafter \@firstoftwo
 \else \expandafter \@secondoftwo
 \fi
}%
\providecommand \natexlab [1]{#1}%
\providecommand \enquote  [1]{``#1''}%
\providecommand \bibnamefont  [1]{#1}%
\providecommand \bibfnamefont [1]{#1}%
\providecommand \citenamefont [1]{#1}%
\providecommand \href@noop [0]{\@secondoftwo}%
\providecommand \href [0]{\begingroup \@sanitize@url \@href}%
\providecommand \@href[1]{\@@startlink{#1}\@@href}%
\providecommand \@@href[1]{\endgroup#1\@@endlink}%
\providecommand \@sanitize@url [0]{\catcode `\\12\catcode `\$12\catcode `\&12\catcode `\#12\catcode `\^12\catcode `\_12\catcode `\%12\relax}%
\providecommand \@@startlink[1]{}%
\providecommand \@@endlink[0]{}%
\providecommand \url  [0]{\begingroup\@sanitize@url \@url }%
\providecommand \@url [1]{\endgroup\@href {#1}{\urlprefix }}%
\providecommand \urlprefix  [0]{URL }%
\providecommand \Eprint [0]{\href }%
\providecommand \doibase [0]{https://doi.org/}%
\providecommand \selectlanguage [0]{\@gobble}%
\providecommand \bibinfo  [0]{\@secondoftwo}%
\providecommand \bibfield  [0]{\@secondoftwo}%
\providecommand \translation [1]{[#1]}%
\providecommand \BibitemOpen [0]{}%
\providecommand \bibitemStop [0]{}%
\providecommand \bibitemNoStop [0]{.\EOS\space}%
\providecommand \EOS [0]{\spacefactor3000\relax}%
\providecommand \BibitemShut  [1]{\csname bibitem#1\endcsname}%
\let\auto@bib@innerbib\@empty
\bibitem [{\citenamefont {Dutra}\ \emph {et~al.}(2014)\citenamefont {Dutra}, \citenamefont {Louren\ifmmode~\mbox{\c{c}}\else \c{c}\fi{}o}, \citenamefont {Avancini} \emph {et~al.}}]{Dutra2014}%
  \BibitemOpen
  \bibfield  {author} {\bibinfo {author} {\bibfnamefont {M.}~\bibnamefont {Dutra}}, \bibinfo {author} {\bibfnamefont {O.}~\bibnamefont {Louren\ifmmode~\mbox{\c{c}}\else \c{c}\fi{}o}}, \bibinfo {author} {\bibfnamefont {S.~S.}\ \bibnamefont {Avancini}}, \emph {et~al.},\ }\bibfield  {title} {\bibinfo {title} {Relativistic mean-field hadronic models under nuclear matter constraints},\ }\href {https://doi.org/10.1103/PhysRevC.90.055203} {\bibfield  {journal} {\bibinfo  {journal} {Phys. Rev. C}\ }\textbf {\bibinfo {volume} {90}},\ \bibinfo {pages} {055203} (\bibinfo {year} {2014})}\BibitemShut {NoStop}%
\bibitem [{\citenamefont {Oertel}\ \emph {et~al.}(2017)\citenamefont {Oertel}, \citenamefont {Hempel}, \citenamefont {Kl\"ahn},\ and\ \citenamefont {Typel}}]{Micaela2017}%
  \BibitemOpen
  \bibfield  {author} {\bibinfo {author} {\bibfnamefont {M.}~\bibnamefont {Oertel}}, \bibinfo {author} {\bibfnamefont {M.}~\bibnamefont {Hempel}}, \bibinfo {author} {\bibfnamefont {T.}~\bibnamefont {Kl\"ahn}},\ and\ \bibinfo {author} {\bibfnamefont {S.}~\bibnamefont {Typel}},\ }\bibfield  {title} {\bibinfo {title} {Equations of state for supernovae and compact stars},\ }\href {https://doi.org/10.1103/RevModPhys.89.015007} {\bibfield  {journal} {\bibinfo  {journal} {Rev. Mod. Phys.}\ }\textbf {\bibinfo {volume} {89}},\ \bibinfo {pages} {015007} (\bibinfo {year} {2017})}\BibitemShut {NoStop}%
\bibitem [{\citenamefont {Paar}\ \emph {et~al.}(2014)\citenamefont {Paar} \emph {et~al.}}]{Paar2014}%
  \BibitemOpen
  \bibfield  {author} {\bibinfo {author} {\bibfnamefont {N.}~\bibnamefont {Paar}} \emph {et~al.},\ }\bibfield  {title} {\bibinfo {title} {Neutron star structure and collective excitations of finite nuclei},\ }\href {https://doi.org/10.1103/PhysRevC.90.011304} {\bibfield  {journal} {\bibinfo  {journal} {Phys. Rev. C}\ }\textbf {\bibinfo {volume} {90}},\ \bibinfo {pages} {011304} (\bibinfo {year} {2014})}\BibitemShut {NoStop}%
\bibitem [{\citenamefont {Lattimer}\ and\ \citenamefont {Steiner}(2014)}]{Steiner2014}%
  \BibitemOpen
  \bibfield  {author} {\bibinfo {author} {\bibfnamefont {J.}~\bibnamefont {Lattimer}}\ and\ \bibinfo {author} {\bibfnamefont {A.}~\bibnamefont {Steiner}},\ }\bibfield  {title} {\bibinfo {title} {Constraints on the symmetry energy using the mass-radius relation of neutron stars},\ }\href {https://doi.org/10.1140/epja/i2014-14040-y} {\bibfield  {journal} {\bibinfo  {journal} {Eur. Phys. J. A}\ }\textbf {\bibinfo {volume} {50}},\ \bibinfo {pages} {40} (\bibinfo {year} {2014})}\BibitemShut {NoStop}%
\bibitem [{\citenamefont {Lattimer}\ and\ \citenamefont {Lim}(2013)}]{Lattimer2013}%
  \BibitemOpen
  \bibfield  {author} {\bibinfo {author} {\bibfnamefont {J.}~\bibnamefont {Lattimer}}\ and\ \bibinfo {author} {\bibfnamefont {Y.}~\bibnamefont {Lim}},\ }\bibfield  {title} {\bibinfo {title} {Constraining the symmetry parameters of the nuclear interaction},\ }\href {https://doi.org/10.1088/0004-637X/771/1/51} {\bibfield  {journal} {\bibinfo  {journal} {Astrophys. J.}\ }\textbf {\bibinfo {volume} {771}},\ \bibinfo {pages} {51} (\bibinfo {year} {2013})}\BibitemShut {NoStop}%
\bibitem [{\citenamefont {Estee}\ \emph {et~al.}(2020)\citenamefont {Estee} \emph {et~al.}}]{pions}%
  \BibitemOpen
  \bibfield  {author} {\bibinfo {author} {\bibfnamefont {J.}~\bibnamefont {Estee}} \emph {et~al.},\ }\bibfield  {title} {\bibinfo {title} {Probing the symmetry energy with the spectral pion ratio},\ }\href {https://doi.org/10.1103/PhysRevLett.126.162701} {\bibfield  {journal} {\bibinfo  {journal} {Phys. Rev. Lett.}\ }\textbf {\bibinfo {volume} {126}},\ \bibinfo {pages} {162701} (\bibinfo {year} {2020})}\BibitemShut {NoStop}%
\bibitem [{\citenamefont {Reed}\ \emph {et~al.}(2020)\citenamefont {Reed} \emph {et~al.}}]{PREX2}%
  \BibitemOpen
  \bibfield  {author} {\bibinfo {author} {\bibfnamefont {B.}~\bibnamefont {Reed}} \emph {et~al.},\ }\bibfield  {title} {\bibinfo {title} {Implications of prex-2 on the equation of state of neutron-rich matter},\ }\href {https://doi.org/10.1103/PhysRevLett.126.172503} {\bibfield  {journal} {\bibinfo  {journal} {Phys. Rev. Lett.}\ }\textbf {\bibinfo {volume} {126}},\ \bibinfo {pages} {172503} (\bibinfo {year} {2020})}\BibitemShut {NoStop}%
\bibitem [{\citenamefont {Tagami}\ \emph {et~al.}(2022)\citenamefont {Tagami}, \citenamefont {Wakasa},\ and\ \citenamefont {Yahiro}}]{Tagami2022}%
  \BibitemOpen
  \bibfield  {author} {\bibinfo {author} {\bibfnamefont {S.}~\bibnamefont {Tagami}}, \bibinfo {author} {\bibfnamefont {T.}~\bibnamefont {Wakasa}},\ and\ \bibinfo {author} {\bibfnamefont {M.}~\bibnamefont {Yahiro}},\ }\bibfield  {title} {\bibinfo {title} {Slope parameters determined from crex and prex2},\ }\href {https://doi.org/https://doi.org/10.1016/j.rinp.2022.106037} {\bibfield  {journal} {\bibinfo  {journal} {Resul. Phys.}\ }\textbf {\bibinfo {volume} {43}},\ \bibinfo {pages} {106037} (\bibinfo {year} {2022})}\BibitemShut {NoStop}%
\bibitem [{\citenamefont {Serot}(1992)}]{Serot_1992}%
  \BibitemOpen
  \bibfield  {author} {\bibinfo {author} {\bibfnamefont {B.~D.}\ \bibnamefont {Serot}},\ }\bibfield  {title} {\bibinfo {title} {Quantum hadrodynamics},\ }\href {https://doi.org/10.1088/0034-4885/55/11/001} {\bibfield  {journal} {\bibinfo  {journal} {Rep. Progr. Phys.}\ }\textbf {\bibinfo {volume} {55}},\ \bibinfo {pages} {1855} (\bibinfo {year} {1992})}\BibitemShut {NoStop}%
\bibitem [{\citenamefont {Miyatsu}\ \emph {et~al.}(2013)\citenamefont {Miyatsu}, \citenamefont {Cheoun},\ and\ \citenamefont {Saito}}]{Miyatsu2013}%
  \BibitemOpen
  \bibfield  {author} {\bibinfo {author} {\bibfnamefont {T.}~\bibnamefont {Miyatsu}}, \bibinfo {author} {\bibfnamefont {M.-K.}\ \bibnamefont {Cheoun}},\ and\ \bibinfo {author} {\bibfnamefont {K.}~\bibnamefont {Saito}},\ }\bibfield  {title} {\bibinfo {title} {Equation of state for neutron stars in su(3) flavor symmetry},\ }\href {https://doi.org/10.1103/PhysRevC.88.015802} {\bibfield  {journal} {\bibinfo  {journal} {Phys. Rev. C}\ }\textbf {\bibinfo {volume} {88}},\ \bibinfo {pages} {015802} (\bibinfo {year} {2013})}\BibitemShut {NoStop}%
\bibitem [{\citenamefont {Antoniadis}\ \emph {et~al.}(2013)\citenamefont {Antoniadis}, \citenamefont {Freire}, \citenamefont {Wex} \emph {et~al.}}]{Antoniadis}%
  \BibitemOpen
  \bibfield  {author} {\bibinfo {author} {\bibfnamefont {J.}~\bibnamefont {Antoniadis}}, \bibinfo {author} {\bibfnamefont {P.~C.~C.}\ \bibnamefont {Freire}}, \bibinfo {author} {\bibfnamefont {N.}~\bibnamefont {Wex}}, \emph {et~al.},\ }\bibfield  {title} {\bibinfo {title} {A massive pulsar in a compact relativistic binary},\ }\href {https://doi.org/10.1126/science.1233232} {\bibfield  {journal} {\bibinfo  {journal} {Science}\ }\textbf {\bibinfo {volume} {340}},\ \bibinfo {pages} {1233232} (\bibinfo {year} {2013})}\BibitemShut {NoStop}%
\bibitem [{\citenamefont {Riley}\ \emph {et~al.}(2021)\citenamefont {Riley} \emph {et~al.}}]{Riley2021}%
  \BibitemOpen
  \bibfield  {author} {\bibinfo {author} {\bibfnamefont {T.}~\bibnamefont {Riley}} \emph {et~al.},\ }\bibfield  {title} {\bibinfo {title} {{A NICER View of the Massive Pulsar PSR J0740+6620 Informed by Radio Timing and XMM-Newton Spectroscopy}},\ }\href {https://doi.org/10.3847/2041-8213/ac0a81} {\bibfield  {journal} {\bibinfo  {journal} {Astrophys. J. Lett.}\ }\textbf {\bibinfo {volume} {918}},\ \bibinfo {pages} {L27} (\bibinfo {year} {2021})}\BibitemShut {NoStop}%
\bibitem [{\citenamefont {Riley}\ \emph {et~al.}(2019)\citenamefont {Riley} \emph {et~al.}}]{Riley:2019yda}%
  \BibitemOpen
  \bibfield  {author} {\bibinfo {author} {\bibfnamefont {T.~E.}\ \bibnamefont {Riley}} \emph {et~al.},\ }\bibfield  {title} {\bibinfo {title} {{A NICER View of PSR J0030+0451: Millisecond Pulsar Parameter Estimation}},\ }\href {https://doi.org/10.3847/2041-8213/ab481c} {\bibfield  {journal} {\bibinfo  {journal} {Astrophys. J. Lett.}\ }\textbf {\bibinfo {volume} {887}},\ \bibinfo {pages} {L21} (\bibinfo {year} {2019})}\BibitemShut {NoStop}%
\bibitem [{\citenamefont {Miller}\ \emph {et~al.}(2019)\citenamefont {Miller} \emph {et~al.}}]{Miller:2019cac}%
  \BibitemOpen
  \bibfield  {author} {\bibinfo {author} {\bibfnamefont {M.}~\bibnamefont {Miller}} \emph {et~al.},\ }\bibfield  {title} {\bibinfo {title} {{PSR J0030+0451 Mass and Radius from $NICER$ Data and Implications for the Properties of Neutron Star Matter}},\ }\href {https://doi.org/10.3847/2041-8213/ab50c5} {\bibfield  {journal} {\bibinfo  {journal} {Astrophys. J. Lett.}\ }\textbf {\bibinfo {volume} {887}},\ \bibinfo {pages} {L24} (\bibinfo {year} {2019})}\BibitemShut {NoStop}%
\bibitem [{\citenamefont {Miller}\ \emph {et~al.}(2021)\citenamefont {Miller} \emph {et~al.}}]{Miller2021}%
  \BibitemOpen
  \bibfield  {author} {\bibinfo {author} {\bibfnamefont {M.}~\bibnamefont {Miller}} \emph {et~al.},\ }\bibfield  {title} {\bibinfo {title} {{The Radius of PSR J0740+6620 from NICER and XMM-Newton Data}},\ }\href {https://doi.org/10.3847/2041-8213/ac089b} {\bibfield  {journal} {\bibinfo  {journal} {Astrophys. J. Lett.}\ }\textbf {\bibinfo {volume} {918}},\ \bibinfo {pages} {L28} (\bibinfo {year} {2021})}\BibitemShut {NoStop}%
\bibitem [{\citenamefont {Annala}\ \emph {et~al.}(2018)\citenamefont {Annala}, \citenamefont {Gorda}, \citenamefont {Kurkela},\ and\ \citenamefont {Vuorinen}}]{Annala2018PRL}%
  \BibitemOpen
  \bibfield  {author} {\bibinfo {author} {\bibfnamefont {E.}~\bibnamefont {Annala}}, \bibinfo {author} {\bibfnamefont {T.}~\bibnamefont {Gorda}}, \bibinfo {author} {\bibfnamefont {A.}~\bibnamefont {Kurkela}},\ and\ \bibinfo {author} {\bibfnamefont {A.}~\bibnamefont {Vuorinen}},\ }\bibfield  {title} {\bibinfo {title} {Gravitational-wave constraints on the neutron-star-matter equation of state},\ }\href {https://doi.org/10.1103/PhysRevLett.120.172703} {\bibfield  {journal} {\bibinfo  {journal} {Phys. Rev. Lett.}\ }\textbf {\bibinfo {volume} {120}},\ \bibinfo {pages} {172703} (\bibinfo {year} {2018})}\BibitemShut {NoStop}%
\bibitem [{\citenamefont {Abbott}\ \emph {et~al.}(2017)\citenamefont {Abbott} \emph {et~al.}}]{Abbott2017}%
  \BibitemOpen
  \bibfield  {author} {\bibinfo {author} {\bibfnamefont {B.}~\bibnamefont {Abbott}} \emph {et~al.},\ }\bibfield  {title} {\bibinfo {title} {Gw170817: Observation of gravitational waves from a binary neutron star inspiral},\ }\href {https://doi.org/10.1103/PhysRevLett.119.161101} {\bibfield  {journal} {\bibinfo  {journal} {Phys. Rev. Lett.}\ }\textbf {\bibinfo {volume} {119}},\ \bibinfo {pages} {161101} (\bibinfo {year} {2017})}\BibitemShut {NoStop}%
\bibitem [{\citenamefont {Abbott}\ \emph {et~al.}(2018)\citenamefont {Abbott}, \citenamefont {Abbott}, \citenamefont {Abbott} \emph {et~al.}}]{AbbottPRL}%
  \BibitemOpen
  \bibfield  {author} {\bibinfo {author} {\bibfnamefont {B.~P.}\ \bibnamefont {Abbott}}, \bibinfo {author} {\bibfnamefont {R.}~\bibnamefont {Abbott}}, \bibinfo {author} {\bibfnamefont {T.~D.}\ \bibnamefont {Abbott}}, \emph {et~al.},\ }\bibfield  {title} {\bibinfo {title} {Gw170817: Measurements of neutron star radii and equation of state},\ }\href {https://doi.org/10.1103/PhysRevLett.121.161101} {\bibfield  {journal} {\bibinfo  {journal} {Phys. Rev. Lett.}\ }\textbf {\bibinfo {volume} {121}},\ \bibinfo {pages} {161101} (\bibinfo {year} {2018})}\BibitemShut {NoStop}%
\bibitem [{\citenamefont {Lattimer}\ \emph {et~al.}(1991)\citenamefont {Lattimer}, \citenamefont {Pethick}, \citenamefont {Prakash},\ and\ \citenamefont {Haensel}}]{Lattimer1991}%
  \BibitemOpen
  \bibfield  {author} {\bibinfo {author} {\bibfnamefont {J.~M.}\ \bibnamefont {Lattimer}}, \bibinfo {author} {\bibfnamefont {C.~J.}\ \bibnamefont {Pethick}}, \bibinfo {author} {\bibfnamefont {M.}~\bibnamefont {Prakash}},\ and\ \bibinfo {author} {\bibfnamefont {P.}~\bibnamefont {Haensel}},\ }\bibfield  {title} {\bibinfo {title} {Direct urca process in neutron stars},\ }\href {https://doi.org/10.1103/PhysRevLett.66.2701} {\bibfield  {journal} {\bibinfo  {journal} {Phys. Rev. Lett.}\ }\textbf {\bibinfo {volume} {66}},\ \bibinfo {pages} {2701} (\bibinfo {year} {1991})}\BibitemShut {NoStop}%
\bibitem [{\citenamefont {Lattimer}\ and\ \citenamefont {Prakash}(2004)}]{Lattimer2004}%
  \BibitemOpen
  \bibfield  {author} {\bibinfo {author} {\bibfnamefont {J.~M.}\ \bibnamefont {Lattimer}}\ and\ \bibinfo {author} {\bibfnamefont {M.}~\bibnamefont {Prakash}},\ }\bibfield  {title} {\bibinfo {title} {The physics of neutron stars},\ }\href {https://doi.org/10.1126/science.1090720} {\bibfield  {journal} {\bibinfo  {journal} {Science}\ }\textbf {\bibinfo {volume} {304}},\ \bibinfo {pages} {536} (\bibinfo {year} {2004})}\BibitemShut {NoStop}%
\bibitem [{\citenamefont {Fattoyev}\ and\ \citenamefont {Piekarewicz}(2012)}]{Fatto2012}%
  \BibitemOpen
  \bibfield  {author} {\bibinfo {author} {\bibfnamefont {F.~J.}\ \bibnamefont {Fattoyev}}\ and\ \bibinfo {author} {\bibfnamefont {J.}~\bibnamefont {Piekarewicz}},\ }\bibfield  {title} {\bibinfo {title} {Neutron skins and neutron stars},\ }\href {https://doi.org/10.1103/PhysRevC.86.015802} {\bibfield  {journal} {\bibinfo  {journal} {Phys. Rev. C}\ }\textbf {\bibinfo {volume} {86}},\ \bibinfo {pages} {015802} (\bibinfo {year} {2012})}\BibitemShut {NoStop}%
\bibitem [{\citenamefont {Dohi}\ \emph {et~al.}(2019)\citenamefont {Dohi} \emph {et~al.}}]{Dohi2019}%
  \BibitemOpen
  \bibfield  {author} {\bibinfo {author} {\bibfnamefont {A.}~\bibnamefont {Dohi}} \emph {et~al.},\ }\bibfield  {title} {\bibinfo {title} {{Possibility of rapid neutron star cooling with a realistic equation of state}},\ }\href {https://doi.org/10.1093/ptep/ptz116} {\bibfield  {journal} {\bibinfo  {journal} {Progr. Theor. Exper. Phys.}\ }\textbf {\bibinfo {volume} {2019}},\ \bibinfo {pages} {113E01} (\bibinfo {year} {2019})}\BibitemShut {NoStop}%
\bibitem [{\citenamefont {Kl\"ahn}\ \emph {et~al.}(2006)\citenamefont {Kl\"ahn} \emph {et~al.}}]{klahn2006}%
  \BibitemOpen
  \bibfield  {author} {\bibinfo {author} {\bibfnamefont {T.}~\bibnamefont {Kl\"ahn}} \emph {et~al.},\ }\bibfield  {title} {\bibinfo {title} {Constraints on the high-density nuclear equation of state from the phenomenology of compact stars and heavy-ion collisions},\ }\href {https://doi.org/10.1103/PhysRevC.74.035802} {\bibfield  {journal} {\bibinfo  {journal} {Phys. Rev. C}\ }\textbf {\bibinfo {volume} {74}},\ \bibinfo {pages} {035802} (\bibinfo {year} {2006})}\BibitemShut {NoStop}%
\bibitem [{\citenamefont {Beznogov}\ and\ \citenamefont {Yakovlev}(2015)}]{Yakovlev2015}%
  \BibitemOpen
  \bibfield  {author} {\bibinfo {author} {\bibfnamefont {M.~V.}\ \bibnamefont {Beznogov}}\ and\ \bibinfo {author} {\bibfnamefont {D.~G.}\ \bibnamefont {Yakovlev}},\ }\bibfield  {title} {\bibinfo {title} {{Statistical theory of thermal evolution of neutron stars – II. Limitations on direct Urca threshold}},\ }\href {https://doi.org/10.1093/mnras/stv1293} {\bibfield  {journal} {\bibinfo  {journal} {MNRAS}\ }\textbf {\bibinfo {volume} {452}},\ \bibinfo {pages} {540} (\bibinfo {year} {2015})}\BibitemShut {NoStop}%
\bibitem [{\citenamefont {{Page}}\ and\ \citenamefont {{Applegate}}(1992)}]{Page1992}%
  \BibitemOpen
  \bibfield  {author} {\bibinfo {author} {\bibfnamefont {D.}~\bibnamefont {{Page}}}\ and\ \bibinfo {author} {\bibfnamefont {J.~H.}\ \bibnamefont {{Applegate}}},\ }\bibfield  {title} {\bibinfo {title} {{The Cooling of Neutron Stars by the Direct URCA Process}},\ }\href {https://doi.org/10.1086/186462} {\bibfield  {journal} {\bibinfo  {journal} {Astrophys. J. Lett.}\ }\textbf {\bibinfo {volume} {394}},\ \bibinfo {pages} {L17} (\bibinfo {year} {1992})}\BibitemShut {NoStop}%
\bibitem [{\citenamefont {Yakovlev}\ \emph {et~al.}(2004)\citenamefont {Yakovlev}, \citenamefont {Gnedin}, \citenamefont {Kaminker}, \citenamefont {Levenfish},\ and\ \citenamefont {Potekhin}}]{YAKOVLEV2004a}%
  \BibitemOpen
  \bibfield  {author} {\bibinfo {author} {\bibfnamefont {D.}~\bibnamefont {Yakovlev}}, \bibinfo {author} {\bibfnamefont {O.}~\bibnamefont {Gnedin}}, \bibinfo {author} {\bibfnamefont {A.}~\bibnamefont {Kaminker}}, \bibinfo {author} {\bibfnamefont {K.}~\bibnamefont {Levenfish}},\ and\ \bibinfo {author} {\bibfnamefont {A.}~\bibnamefont {Potekhin}},\ }\bibfield  {title} {\bibinfo {title} {Neutron star cooling: theoretical aspects and observational constraints},\ }\href {https://doi.org/https://doi.org/10.1016/j.asr.2003.07.020} {\bibfield  {journal} {\bibinfo  {journal} {Advan. Spac. Res.}\ }\textbf {\bibinfo {volume} {33}},\ \bibinfo {pages} {523} (\bibinfo {year} {2004})}\BibitemShut {NoStop}%
\bibitem [{\citenamefont {Yakovlev}\ and\ \citenamefont {Pethick}(2004)}]{Yakovlev2004b}%
  \BibitemOpen
  \bibfield  {author} {\bibinfo {author} {\bibfnamefont {D.}~\bibnamefont {Yakovlev}}\ and\ \bibinfo {author} {\bibfnamefont {C.}~\bibnamefont {Pethick}},\ }\bibfield  {title} {\bibinfo {title} {Neutron star cooling},\ }\href {https://doi.org/10.1146/annurev.astro.42.053102.134013} {\bibfield  {journal} {\bibinfo  {journal} {Ann. Rev. Astron. Astrophys.}\ }\textbf {\bibinfo {volume} {42}},\ \bibinfo {pages} {169} (\bibinfo {year} {2004})}\BibitemShut {NoStop}%
\bibitem [{\citenamefont {Fattoyev}\ \emph {et~al.}(2010)\citenamefont {Fattoyev} \emph {et~al.}}]{IUFSU}%
  \BibitemOpen
  \bibfield  {author} {\bibinfo {author} {\bibfnamefont {F.}~\bibnamefont {Fattoyev}} \emph {et~al.},\ }\bibfield  {title} {\bibinfo {title} {Relativistic effective interaction for nuclei, giant resonances, and neutron stars},\ }\href {https://doi.org/10.1103/PhysRevC.82.055803} {\bibfield  {journal} {\bibinfo  {journal} {Phys. Rev. C}\ }\textbf {\bibinfo {volume} {82}},\ \bibinfo {pages} {055803} (\bibinfo {year} {2010})}\BibitemShut {NoStop}%
\bibitem [{\citenamefont {Cavagnoli}\ \emph {et~al.}(2011)\citenamefont {Cavagnoli}, \citenamefont {Menezes},\ and\ \citenamefont {Providencias}}]{Rafa2011}%
  \BibitemOpen
  \bibfield  {author} {\bibinfo {author} {\bibfnamefont {R.}~\bibnamefont {Cavagnoli}}, \bibinfo {author} {\bibfnamefont {D.}~\bibnamefont {Menezes}},\ and\ \bibinfo {author} {\bibfnamefont {C.}~\bibnamefont {Providencias}},\ }\bibfield  {title} {\bibinfo {title} {Neutron star properties and the symmetry energy},\ }\href {https://doi.org/10.1103/PhysRevC.84.065810} {\bibfield  {journal} {\bibinfo  {journal} {Phys. Rev. C}\ }\textbf {\bibinfo {volume} {84}},\ \bibinfo {pages} {065810} (\bibinfo {year} {2011})}\BibitemShut {NoStop}%
\bibitem [{\citenamefont {Dexheimer}\ \emph {et~al.}(2019)\citenamefont {Dexheimer} \emph {et~al.}}]{dex19jpg}%
  \BibitemOpen
  \bibfield  {author} {\bibinfo {author} {\bibfnamefont {V.}~\bibnamefont {Dexheimer}} \emph {et~al.},\ }\bibfield  {title} {\bibinfo {title} {What do we learn about vector interactions from gw170817?},\ }\href {https://doi.org/10.1088/1361-6471/ab01f0} {\bibfield  {journal} {\bibinfo  {journal} {J. Phys. G}\ }\textbf {\bibinfo {volume} {46}},\ \bibinfo {pages} {034002} (\bibinfo {year} {2019})}\BibitemShut {NoStop}%
\bibitem [{\citenamefont {Kubis}\ and\ \citenamefont {Kutschera}(1997)}]{KUBIS1997}%
  \BibitemOpen
  \bibfield  {author} {\bibinfo {author} {\bibfnamefont {S.}~\bibnamefont {Kubis}}\ and\ \bibinfo {author} {\bibfnamefont {M.}~\bibnamefont {Kutschera}},\ }\bibfield  {title} {\bibinfo {title} {Nuclear matter in relativistic mean field theory with isovector scalar meson},\ }\href {https://doi.org/https://doi.org/10.1016/S0370-2693(97)00306-7} {\bibfield  {journal} {\bibinfo  {journal} {Phys. Lett. B}\ }\textbf {\bibinfo {volume} {399}},\ \bibinfo {pages} {191} (\bibinfo {year} {1997})}\BibitemShut {NoStop}%
\bibitem [{\citenamefont {Liu}\ \emph {et~al.}(2002)\citenamefont {Liu}, \citenamefont {Greco}, \citenamefont {Baran}, \citenamefont {Colonna},\ and\ \citenamefont {Di~Toro}}]{Liu2002}%
  \BibitemOpen
  \bibfield  {author} {\bibinfo {author} {\bibfnamefont {B.}~\bibnamefont {Liu}}, \bibinfo {author} {\bibfnamefont {V.}~\bibnamefont {Greco}}, \bibinfo {author} {\bibfnamefont {V.}~\bibnamefont {Baran}}, \bibinfo {author} {\bibfnamefont {M.}~\bibnamefont {Colonna}},\ and\ \bibinfo {author} {\bibfnamefont {M.}~\bibnamefont {Di~Toro}},\ }\bibfield  {title} {\bibinfo {title} {Asymmetric nuclear matter: The role of the isovector scalar channel},\ }\href {https://doi.org/10.1103/PhysRevC.65.045201} {\bibfield  {journal} {\bibinfo  {journal} {Phys. Rev. C}\ }\textbf {\bibinfo {volume} {65}},\ \bibinfo {pages} {045201} (\bibinfo {year} {2002})}\BibitemShut {NoStop}%
\bibitem [{\citenamefont {Lopes}\ and\ \citenamefont {Menezes}(2014)}]{Lopes2014BJP}%
  \BibitemOpen
  \bibfield  {author} {\bibinfo {author} {\bibfnamefont {L.}~\bibnamefont {Lopes}}\ and\ \bibinfo {author} {\bibfnamefont {D.}~\bibnamefont {Menezes}},\ }\bibfield  {title} {\bibinfo {title} {Effects of the symmetry energy and its slope on neutron star properties},\ }\href {https://doi.org/10.1007/s13538-014-0252-4} {\bibfield  {journal} {\bibinfo  {journal} {Braz. J. Phys.}\ }\textbf {\bibinfo {volume} {44}},\ \bibinfo {pages} {774} (\bibinfo {year} {2014})}\BibitemShut {NoStop}%
\bibitem [{\citenamefont {Lopes}\ \emph {et~al.}(2023)\citenamefont {Lopes} \emph {et~al.}}]{lopescesar}%
  \BibitemOpen
  \bibfield  {author} {\bibinfo {author} {\bibfnamefont {L.~L.}\ \bibnamefont {Lopes}} \emph {et~al.},\ }\bibfield  {title} {\bibinfo {title} {Imprints of the nuclear symmetry energy slope in gravitational wave signals emanating from neutron stars},\ }\href {https://doi.org/10.1103/PhysRevD.108.083042} {\bibfield  {journal} {\bibinfo  {journal} {Phys. Rev. D}\ }\textbf {\bibinfo {volume} {108}},\ \bibinfo {pages} {083042} (\bibinfo {year} {2023})}\BibitemShut {NoStop}%
\bibitem [{\citenamefont {Lopes}\ \emph {et~al.}(2024)\citenamefont {Lopes}, \citenamefont {Menezes},\ and\ \citenamefont {Pelicer}}]{lopes2024PRC}%
  \BibitemOpen
  \bibfield  {author} {\bibinfo {author} {\bibfnamefont {L.~L.}\ \bibnamefont {Lopes}}, \bibinfo {author} {\bibfnamefont {D.~P.}\ \bibnamefont {Menezes}},\ and\ \bibinfo {author} {\bibfnamefont {M.~R.}\ \bibnamefont {Pelicer}},\ }\bibfield  {title} {\bibinfo {title} {Correlation between the symmetry energy slope and the deconfinement phase transition},\ }\href {https://doi.org/10.1103/PhysRevC.109.045801} {\bibfield  {journal} {\bibinfo  {journal} {Phys. Rev. C}\ }\textbf {\bibinfo {volume} {109}},\ \bibinfo {pages} {045801} (\bibinfo {year} {2024})}\BibitemShut {NoStop}%
\bibitem [{\citenamefont {Boguta}\ and\ \citenamefont {Bodmer}(1977)}]{Boguta}%
  \BibitemOpen
  \bibfield  {author} {\bibinfo {author} {\bibfnamefont {J.}~\bibnamefont {Boguta}}\ and\ \bibinfo {author} {\bibfnamefont {A.}~\bibnamefont {Bodmer}},\ }\bibfield  {title} {\bibinfo {title} {Relativistic calculation of nuclear matter and the nuclear surface},\ }\href {https://doi.org/10.1016/0375-9474(77)90626-1} {\bibfield  {journal} {\bibinfo  {journal} {Nucl. Phys. A}\ }\textbf {\bibinfo {volume} {292}},\ \bibinfo {pages} {413} (\bibinfo {year} {1977})}\BibitemShut {NoStop}%
\bibitem [{\citenamefont {Lopes}\ and\ \citenamefont {Menezes}(2022)}]{Lopes2022ApJ}%
  \BibitemOpen
  \bibfield  {author} {\bibinfo {author} {\bibfnamefont {L.~L.}\ \bibnamefont {Lopes}}\ and\ \bibinfo {author} {\bibfnamefont {D.~P.}\ \bibnamefont {Menezes}},\ }\bibfield  {title} {\bibinfo {title} {On the nature of the mass-gap object in the gw190814 event},\ }\href {https://doi.org/10.3847/1538-4357/ac81c4} {\bibfield  {journal} {\bibinfo  {journal} {Astrophys. J.}\ }\textbf {\bibinfo {volume} {936}},\ \bibinfo {pages} {41} (\bibinfo {year} {2022})}\BibitemShut {NoStop}%
\bibitem [{\citenamefont {Tolos}\ \emph {et~al.}(2017)\citenamefont {Tolos}, \citenamefont {Centelles},\ and\ \citenamefont {Ramos}}]{Tolos2017}%
  \BibitemOpen
  \bibfield  {author} {\bibinfo {author} {\bibfnamefont {L.}~\bibnamefont {Tolos}}, \bibinfo {author} {\bibfnamefont {M.}~\bibnamefont {Centelles}},\ and\ \bibinfo {author} {\bibfnamefont {A.}~\bibnamefont {Ramos}},\ }\bibfield  {title} {\bibinfo {title} {The equation of state for the nucleonic and hyperonic core of neutron stars},\ }\href {https://doi.org/10.1017/pasa.2017.60} {\bibfield  {journal} {\bibinfo  {journal} {PASA}\ }\textbf {\bibinfo {volume} {34}},\ \bibinfo {pages} {E065} (\bibinfo {year} {2017})}\BibitemShut {NoStop}%
\bibitem [{\citenamefont {Glendenning}(2000)}]{Glenbook}%
  \BibitemOpen
  \bibfield  {author} {\bibinfo {author} {\bibfnamefont {N.~K.}\ \bibnamefont {Glendenning}},\ }\href@noop {} {\emph {\bibinfo {title} {Compact stars:}}}\ (\bibinfo  {publisher} {2 ed. Edition, Springer New York},\ \bibinfo {year} {2000})\BibitemShut {NoStop}%
\bibitem [{\citenamefont {Lopes}(2022)}]{Lopes2022CTP}%
  \BibitemOpen
  \bibfield  {author} {\bibinfo {author} {\bibfnamefont {L.~L.}\ \bibnamefont {Lopes}},\ }\bibfield  {title} {\bibinfo {title} {Hyperonic neutron stars: reconciliation between nuclear properties and nicer and ligo/virgo results},\ }\href {https://doi.org/10.1088/1572-9494/ac2297} {\bibfield  {journal} {\bibinfo  {journal} {Commun. Theor. Phys.}\ }\textbf {\bibinfo {volume} {74}},\ \bibinfo {pages} {015302} (\bibinfo {year} {2022})}\BibitemShut {NoStop}%
\bibitem [{\citenamefont {Glendenning}\ and\ \citenamefont {Moszkowski}(1991)}]{GlenPRL}%
  \BibitemOpen
  \bibfield  {author} {\bibinfo {author} {\bibfnamefont {N.~K.}\ \bibnamefont {Glendenning}}\ and\ \bibinfo {author} {\bibfnamefont {S.~A.}\ \bibnamefont {Moszkowski}},\ }\bibfield  {title} {\bibinfo {title} {Reconciliation of neutron-star masses and binding of the \ensuremath{\Lambda} in hypernuclei},\ }\href {https://doi.org/10.1103/PhysRevLett.67.2414} {\bibfield  {journal} {\bibinfo  {journal} {Phys. Rev. Lett.}\ }\textbf {\bibinfo {volume} {67}},\ \bibinfo {pages} {2414} (\bibinfo {year} {1991})}\BibitemShut {NoStop}%
\bibitem [{\citenamefont {Fattoyev}\ \emph {et~al.}(2020)\citenamefont {Fattoyev}, \citenamefont {Horowitz}, \citenamefont {Piekarewicz},\ and\ \citenamefont {Reed}}]{BigApple}%
  \BibitemOpen
  \bibfield  {author} {\bibinfo {author} {\bibfnamefont {F.~J.}\ \bibnamefont {Fattoyev}}, \bibinfo {author} {\bibfnamefont {C.~J.}\ \bibnamefont {Horowitz}}, \bibinfo {author} {\bibfnamefont {J.}~\bibnamefont {Piekarewicz}},\ and\ \bibinfo {author} {\bibfnamefont {B.}~\bibnamefont {Reed}},\ }\bibfield  {title} {\bibinfo {title} {Gw190814: Impact of a 2.6 solar mass neutron star on the nucleonic equations of state},\ }\href {https://doi.org/10.1103/PhysRevC.102.065805} {\bibfield  {journal} {\bibinfo  {journal} {Phys. Rev. C}\ }\textbf {\bibinfo {volume} {102}},\ \bibinfo {pages} {065805} (\bibinfo {year} {2020})}\BibitemShut {NoStop}%
\bibitem [{\citenamefont {Malik}\ \emph {et~al.}(2024)\citenamefont {Malik}, \citenamefont {Dexheimer},\ and\ \citenamefont {Providência}}]{Dex2024arxiv}%
  \BibitemOpen
  \bibfield  {author} {\bibinfo {author} {\bibfnamefont {T.}~\bibnamefont {Malik}}, \bibinfo {author} {\bibfnamefont {V.}~\bibnamefont {Dexheimer}},\ and\ \bibinfo {author} {\bibfnamefont {C.}~\bibnamefont {Providência}},\ }\bibfield  {title} {\bibinfo {title} {Astrophysics and nuclear physics informed interactions in dense matter: Insights from psr j0437-4715},\ }\bibfield  {journal} {\bibinfo  {journal} {arXiv}\ }\href {https://doi.org/10.48550/arXiv.2404.07936} {10.48550/arXiv.2404.07936} (\bibinfo {year} {2024})\BibitemShut {NoStop}%
\bibitem [{\citenamefont {Romani}\ \emph {et~al.}(2022)\citenamefont {Romani}, \citenamefont {Kandel}, \citenamefont {Filippenko}, \citenamefont {Brink},\ and\ \citenamefont {Zheng}}]{romani}%
  \BibitemOpen
  \bibfield  {author} {\bibinfo {author} {\bibfnamefont {R.~W.}\ \bibnamefont {Romani}}, \bibinfo {author} {\bibfnamefont {D.}~\bibnamefont {Kandel}}, \bibinfo {author} {\bibfnamefont {A.~V.}\ \bibnamefont {Filippenko}}, \bibinfo {author} {\bibfnamefont {T.~G.}\ \bibnamefont {Brink}},\ and\ \bibinfo {author} {\bibfnamefont {W.}~\bibnamefont {Zheng}},\ }\bibfield  {title} {\bibinfo {title} {Psr j0952-0607: The fastest and heaviest known galactic neutron star},\ }\href {https://doi.org/10.3847/2041-8213/ac8007} {\bibfield  {journal} {\bibinfo  {journal} {Astrophys. J. Lett.}\ }\textbf {\bibinfo {volume} {934}},\ \bibinfo {pages} {L17} (\bibinfo {year} {2022})}\BibitemShut {NoStop}%
\bibitem [{\citenamefont {Oppenheimer}\ and\ \citenamefont {Volkoff}(1939)}]{TOV}%
  \BibitemOpen
  \bibfield  {author} {\bibinfo {author} {\bibfnamefont {J.~R.}\ \bibnamefont {Oppenheimer}}\ and\ \bibinfo {author} {\bibfnamefont {G.~M.}\ \bibnamefont {Volkoff}},\ }\bibfield  {title} {\bibinfo {title} {On massive neutron cores},\ }\href {https://doi.org/10.1103/PhysRev.55.374} {\bibfield  {journal} {\bibinfo  {journal} {Phys. Rev.}\ }\textbf {\bibinfo {volume} {55}},\ \bibinfo {pages} {374} (\bibinfo {year} {1939})}\BibitemShut {NoStop}%
\bibitem [{\citenamefont {Baym}\ \emph {et~al.}(1971{\natexlab{a}})\citenamefont {Baym}, \citenamefont {Pethick},\ and\ \citenamefont {Sutherland}}]{BPS}%
  \BibitemOpen
  \bibfield  {author} {\bibinfo {author} {\bibfnamefont {G.}~\bibnamefont {Baym}}, \bibinfo {author} {\bibfnamefont {C.}~\bibnamefont {Pethick}},\ and\ \bibinfo {author} {\bibfnamefont {P.}~\bibnamefont {Sutherland}},\ }\bibfield  {title} {\bibinfo {title} {The ground state of matter at high densities},\ }\href {https://doi.org/10.1086/151216} {\bibfield  {journal} {\bibinfo  {journal} {Astrophys. J.}\ }\textbf {\bibinfo {volume} {170}},\ \bibinfo {pages} {299} (\bibinfo {year} {1971}{\natexlab{a}})}\BibitemShut {NoStop}%
\bibitem [{\citenamefont {Baym}\ \emph {et~al.}(1971{\natexlab{b}})\citenamefont {Baym}, \citenamefont {Bethe},\ and\ \citenamefont {Pethick}}]{BBP}%
  \BibitemOpen
  \bibfield  {author} {\bibinfo {author} {\bibfnamefont {G.}~\bibnamefont {Baym}}, \bibinfo {author} {\bibfnamefont {H.~A.}\ \bibnamefont {Bethe}},\ and\ \bibinfo {author} {\bibfnamefont {C.~J.}\ \bibnamefont {Pethick}},\ }\bibfield  {title} {\bibinfo {title} {Neutron star matter},\ }\href {https://doi.org/https://doi.org/10.1016/0375-9474(71)90281-8} {\bibfield  {journal} {\bibinfo  {journal} {Nucl.Phys. A}\ }\textbf {\bibinfo {volume} {175}},\ \bibinfo {pages} {225} (\bibinfo {year} {1971}{\natexlab{b}})}\BibitemShut {NoStop}%
\bibitem [{\citenamefont {Fortin}\ \emph {et~al.}(2016)\citenamefont {Fortin} \emph {et~al.}}]{Fortin2016}%
  \BibitemOpen
  \bibfield  {author} {\bibinfo {author} {\bibfnamefont {M.}~\bibnamefont {Fortin}} \emph {et~al.},\ }\bibfield  {title} {\bibinfo {title} {Neutron star radii and crusts: Uncertainties and unified equations of state},\ }\href {https://doi.org/10.1103/PhysRevC.94.035804} {\bibfield  {journal} {\bibinfo  {journal} {Phys. Rev. C}\ }\textbf {\bibinfo {volume} {94}},\ \bibinfo {pages} {035804} (\bibinfo {year} {2016})}\BibitemShut {NoStop}%
\bibitem [{\citenamefont {Lopes}(2024)}]{Lopes2024ApJ}%
  \BibitemOpen
  \bibfield  {author} {\bibinfo {author} {\bibfnamefont {L.~L.}\ \bibnamefont {Lopes}},\ }\bibfield  {title} {\bibinfo {title} {Decoding rotating neutron stars: Role of the symmetry energy slope},\ }\href {https://doi.org/10.3847/1538-4357/ad391e} {\bibfield  {journal} {\bibinfo  {journal} {Astrophys. J.}\ }\textbf {\bibinfo {volume} {966}},\ \bibinfo {pages} {184} (\bibinfo {year} {2024})}\BibitemShut {NoStop}%
\bibitem [{\citenamefont {Flanagan}\ and\ \citenamefont {Hinderer}(2008)}]{Eanna2008}%
  \BibitemOpen
  \bibfield  {author} {\bibinfo {author} {\bibfnamefont {E.~E.}\ \bibnamefont {Flanagan}}\ and\ \bibinfo {author} {\bibfnamefont {T.}~\bibnamefont {Hinderer}},\ }\bibfield  {title} {\bibinfo {title} {Constraining neutron-star tidal love numbers with gravitational-wave detectors},\ }\href {https://doi.org/10.1103/PhysRevD.77.021502} {\bibfield  {journal} {\bibinfo  {journal} {Phys. Rev. D}\ }\textbf {\bibinfo {volume} {77}},\ \bibinfo {pages} {021502} (\bibinfo {year} {2008})}\BibitemShut {NoStop}%
\bibitem [{\citenamefont {Hinderer}(2008)}]{Hinderer_2008}%
  \BibitemOpen
  \bibfield  {author} {\bibinfo {author} {\bibfnamefont {T.}~\bibnamefont {Hinderer}},\ }\bibfield  {title} {\bibinfo {title} {Tidal love numbers of neutron stars},\ }\href {https://doi.org/10.1086/533487} {\bibfield  {journal} {\bibinfo  {journal} {Astrophys. J.}\ }\textbf {\bibinfo {volume} {677}},\ \bibinfo {pages} {1216} (\bibinfo {year} {2008})}\BibitemShut {NoStop}%
\bibitem [{\citenamefont {Chatziioannou}(2020)}]{Chat2020}%
  \BibitemOpen
  \bibfield  {author} {\bibinfo {author} {\bibfnamefont {K.}~\bibnamefont {Chatziioannou}},\ }\href {https://doi.org/10.1007/s10714-020-02754-3} {\bibfield  {journal} {\bibinfo  {journal} {Gen. Rel. Grav}\ }\textbf {\bibinfo {volume} {52}},\ \bibinfo {pages} {109} (\bibinfo {year} {2020})}\BibitemShut {NoStop}%
\bibitem [{\citenamefont {Flores}\ \emph {et~al.}(2020)\citenamefont {Flores}, \citenamefont {Lopes}, \citenamefont {Benito},\ and\ \citenamefont {Menezes}}]{Flores2020}%
  \BibitemOpen
  \bibfield  {author} {\bibinfo {author} {\bibfnamefont {C.}~\bibnamefont {Flores}}, \bibinfo {author} {\bibfnamefont {L.}~\bibnamefont {Lopes}}, \bibinfo {author} {\bibfnamefont {L.}~\bibnamefont {Benito}},\ and\ \bibinfo {author} {\bibfnamefont {D.}~\bibnamefont {Menezes}},\ }\bibfield  {title} {\bibinfo {title} {Gravitational wave signatures of highly magnetized neutron stars},\ }\href {https://doi.org/10.1140/epjc/s10052-020-08705-1} {\bibfield  {journal} {\bibinfo  {journal} {Eur. Phys. J. C}\ }\textbf {\bibinfo {volume} {80}},\ \bibinfo {pages} {1142} (\bibinfo {year} {2020})}\BibitemShut {NoStop}%
\end{thebibliography}%

\end{document}